\documentclass[peerreview]{IEEEtran}

\usepackage{epsf}
\usepackage{amssymb}
\usepackage{graphicx}

\newtheorem{thm}{Theorem}
\newtheorem{lemma}[thm]{Lemma}
\newtheorem{definition}[thm]{Definition}
\newtheorem{proposition}[thm]{Proposition}
\newtheorem{corollary}[thm]{Corollary}

\begin{document}

\title {Guessing Under Source Uncertainty}

\author {Rajesh~Sundaresan,~\IEEEmembership{Senior Member,~IEEE}
\thanks{R. Sundaresan is with the Department of Electrical Communication
Engineering, Indian Institute of Science, Bangalore 560012, India}
\thanks{This work was supported in part by the Ministry of Human Resources
and Development under Grant Part (2A) Tenth Plan (338/ECE), by the
DRDO-IISc joint programme on mathematical engineering, and the
University Grants Commission under Grant Part (2B) UGC-CAS-(Ph.IV).
Material in this paper was presented in part at the IEEE
International Symposium on Information Theory (ISIT 2002) held in
Lausanne, Switzerland, July 2002, at the National Conference on
Communication (NCC 2006), New Delhi, India, Jan. 2006, at the
Conference on Information Sciences and Systems (CISS 2006),
Princeton, NJ, March, 2006, and at IEEE International Symposium on
Information Theory (ISIT 2006) held in Seattle, WA, USA, July 2006.
} }

\maketitle

\begin{abstract}

This paper considers the problem of guessing the realization of a
finite alphabet source when some side information is provided. The
only knowledge the guesser has about the source and the correlated
side information is that the joint source is one among a family. A
notion of redundancy is first defined and a new divergence quantity
that measures this redundancy is identified. This divergence
quantity shares the Pythagorean property with the Kullback-Leibler
divergence. Good guessing strategies that minimize the supremum
redundancy (over the family) are then identified. The min-sup value
measures the richness of the uncertainty set. The min-sup
redundancies for two examples - the families of discrete memoryless
sources and finite-state arbitrarily varying sources - are then
determined.

\end{abstract}


\section{Introduction}

Let $X$ be a random variable on a finite set $\mathbb{X}$ with
probability mass function (PMF) given by $(P(x): x \in \mathbb{X})$.
Suppose that we wish to guess the realization of this random
variable $X$ by asking questions of the form ``Is $X$ equal to
$x$?'', stepping through the elements of $\mathbb{X}$, until the
answer is ``Yes'' (\cite{1994_ProcISIT_Mas}, \cite{Arikan}). If we
know the PMF $P$, the best strategy is to guess in the decreasing
order of $P$-probabilities.

The aim of this paper is to identify good guessing strategies and
analyze their performance when the PMF $P$ is not completely known.
Throughout this paper, we will assume that the only information
available to the guesser is that the PMF of the source is one among
a family $\mathbb{T}$ of PMFs.

By way of motivation, consider a crypto-system in which Alice wishes
to send a secret message to Bob. The message is encrypted using a
private key stream. Alice and Bob share this private key stream. The
key stream is generated using a random and perhaps biased source.
The cipher-text is transmitted through a public channel. Eve, the
eavesdropper, guesses one key stream after another until she arrives
at the correct message. Eve can guess any number of times, and she
knows when she has guessed right. She might know this, for example,
when she obtains a meaningful message. From Alice's and Bob's points
of view, how good is their key stream generating source? In
particular, what is the minimum expected number of guesses that Eve
would need to get to the correct realization? From Eve's point of
view, what is her best guessing strategy? These questions were
answered by Arikan in \cite{Arikan} and generalized to systems with
specified key rate by Merhav and Arikan in \cite{199909TIT_MerAri}.

Taking this example a step further, suppose that Alice and Bob have
access to a few sources. How can they utilize these sources to
increase the expected number of guesses Eve will need? What is Eve's
guessing strategy? We answer these questions in this paper.

When $P$ is known, Massey \cite{1994_ProcISIT_Mas} and Arikan
\cite{Arikan} sought to lower bound the minimum expected number of
guesses. For a given guessing strategy $G$, let $G(x)$ denote the
number of guesses required when $X=x$. The strategy that minimizes
$\mathbb{E}\left[G(X)\right]$, the expected number of guesses,
proceeds in the decreasing order of $P$-probabilities. Arikan
\cite{Arikan} showed that the exponent of the minimum value, {\it
i.e.}, $ \log \left[ \min_{G} ~ \mathbb{E} \left[ G(X) \right]
\right] $, satisfies
\[
H_{1/2}(P) - \log(1+\ln|{\mathbb{X}}|) \leq \log \left[ \min_{G} ~
\mathbb{E} \left[ G(X) \right] \right] \leq H_{1/2}(P),
\]
where $H_{\alpha}(P)$ is the R\'{e}nyi entropy of order ${\alpha} >
0$. Bozta\c{s} \cite{199711TIT_Boz} obtains a tighter upper bound.

For $\rho > 0$, Arikan \cite{Arikan} also considered minimization of
$\left(\mathbb{E}[G(X)^{\rho}]\right)^{1/\rho}$ over all guessing
strategies $G$; the exponent of the minimum value satisfies
\begin{equation}
\label{ExponentInequality} H_{\alpha}(P) - \log(1+\ln|{\mathbb{X}}|)
\leq \frac{1}{\rho} \log \left[ \min_{G} \mathbb{E} \left[
G(X)^{\rho} \right] \right] \leq H_{\alpha}(P),
\end{equation}
where $\alpha = 1/(1+\rho)$.

Arikan \cite{Arikan} applied these results to a discrete memoryless
source on $\mathbb{X}$ with letter probabilities given by the PMF
$P$, and obtained that the minimum guessing moment, $\min_{G}
\mathbb{E} \left[ G(X^n)^{\rho} \right]$, grows exponentially with
$n$. The minimum growth rate of this quantity (after normalization
by $\rho$) is given by the R\'{e}nyi entropy $H_{\alpha}(P)$. This
gave an operational significance for the R\'{e}nyi entropy. In
particular, the minimum expected number of guesses grows
exponentially with $n$ and has a minimum growth rate of
$H_{1/2}(P)$. The study of $\mathbb{E} \left[ G(X)^{\rho}\right]$,
as a function of $\rho$, is motivated by the fact that it is the
convex conjugate (Legendre-Fenchel transformation) of a function
that characterizes the large deviations behavior of the number of
guesses. See \cite{199909TIT_MerAri} for more details.

Suppose now that the guesser only knows that the source belongs to a
family $\mathbb{T}$ of PMFs. The uncertainty set may be finite or
infinite in size. The guesser's strategy should not be tuned to any
one particular PMF in $\mathbb{T}$, but should be designed for the
entire uncertainty set. The performance of such a guessing strategy
on any particular source will not be better than the optimal
strategy for that source. Indeed, for any source $P$, the exponent
of $\mathbb{E} \left[ G(X)^{\rho} \right]$ is at least as large as
that of the optimal strategy $\mathbb{E} \left[ G_P(X)^{\rho}
\right]$, where $G_P$ is the guessing strategy matched to $P$ that
guesses in the decreasing order of $P$-probabilities. Thus for any
given strategy, and for any source $P \in \mathbb{T}$, we can define
a notion of {\em penalty} or {\em redundancy}, $R(P,G)$, given by
\[
R(P,G) = \frac{1}{\rho} \log \mathbb{E} \left[ G(X)^{\rho} \right] -
\frac{1}{\rho} \log  \mathbb{E} \left[ G_P(X)^{\rho} \right],
\]
which represents the increase in the exponent of the guessing moment
normalized by $\rho$.

A natural means of measuring the effectiveness of a guessing
strategy $G$ on the family $\mathbb{T}$ is to find the worst
redundancy over all sources in $\mathbb{T}$. In this paper, we are
interested in identifying the value of
\[
\min_G \sup_{P \in \mathbb{T}} R(P,G),
\]
and in obtaining the $G$ that attains this min-sup value.

We first show that $R(P,G)$ is bounded on either side in terms of a
divergence quantity $L_{\alpha}(P, Q_G)$; $Q_G$ is a PMF that
depends on $G$, and $L_{\alpha}$ is a measure of dissimilarity
between two PMFs. The above observation enables us to transform the
min-sup problem above into another one of identifying
\[
\inf_Q \sup_{P \in \mathbb{T}} L_{\alpha}(P,Q).
\]
The role of $L_{\alpha}$ in guessing is similar to the role of
Kullback-Leibler divergence in mismatched source compression. The
parameter $\alpha$ is given by $\alpha = 1/(1+\rho)$. The quantity
$L_{\alpha}$ is such that the limiting value as $\alpha \rightarrow
1$ is the Kullback-Leibler divergence. Furthermore, $L_{\alpha}$
shares the Pythagorean property with the Kullback-Leibler divergence
\cite{Csiszar-I-proj}. The results of this paper thus generalize the
``geometric'' properties satisfied by the Kullback-Leibler
divergence \cite{Csiszar-I-proj}.

Consider the special case of guessing an $n$-string put out by a
discrete memoryless source (DMS) with single letter alphabet
$\mathbb{A}$. The parameters of this DMS are unknown to the guesser.
Arikan and Merhav \cite{Arikan-Merhav} proposed a ``universal''
guessing strategy for the family of DMSs on $\mathbb{A}$. This
universal guessing strategy asymptotically achieves the minimum
growth exponent for all sources in the uncertainty set. Their
strategy guesses in the increasing order of empirical entropy. In
the language of this paper, their results imply that the normalized
redundancy suffered by the aforementioned strategy is upper bounded
by a positive sequence of real numbers that vanishes as $n
\rightarrow \infty$. One can interpret this fact as follows: the
family of discrete memoryless sources is not ``rich'' enough; we
have a universal guessing strategy that is asymptotically optimal.

The redundancy quantities studied in this paper also arise in the
study of mismatch in Campbell's minimum average exponential coding
length problem. Campbell (\cite{Campbell-1} and \cite{Campbell-2})
identified a code that depended on knowledge of the source PMF. The
code has redundancy within a constant of the optimal value and is
analogous to the Shannon code for source compression. Blumer and
McEliece \cite{198809TIT_BluMcE} studied a modified Huffman
algorithm for this problem and tightened the bounds on the
redundancy. Fischer \cite{Fis-Prag78} addressed the problem in the
context of mismatched source compression and identified the supremum
average exponential coding length for a family of sources. In
particular, he showed that the supremum value is the supremum of the
R\'{e}nyi entropies of the sources in the family. In contrast to
Fischer's work, our focus in this paper is on identifying the worst
{\it redundancy} suffered by a code.

Most of the results obtained in this paper were inspired by similar
results for mismatched and universal source compression
(\cite{197211TIT_Dav}, \cite{197909TR_Gal},
\cite{Davisson-Leongarcia}). We now highlight some comparisons
between source compression and guessing.

Suppose that the source outputs an $n$-string of bits. In lossless
source compression, one can think of an encoding scheme as asking
questions of the form, ``Does $X^n \in E_i$?'' where $(E_i : i = 1,
2, \cdots)$ is a carefully chosen sequence of subsets of
$\mathbb{X}^n$. More specifically, one can ask the questions ``Is
$X_1 = 0$?'', ``Is $X_2 = 0$?'', and so on. The goal is to minimize
the number of such questions one needs to ask (on the average) to
get to the realization. The minimum expected number of questions one
can hope to ask (on the average) is the Shannon entropy $H(P)$. In
the context of guessing, one can only test an entire string in one
attempt, {\it i.e.}, ask questions of the form ``Is $X^n = x^n$?''.
The guessing moment grows exponentially with $n$ and the minimum
exponent, after scaling by $\rho$, is given by the R\'{e}nyi entropy
$H_{\alpha}(P)$.

The quantity $L_{\alpha}$ plays the same role as Kullback-Leibler
divergence does in mismatched source compression. $L_{\alpha}$
shares the Pythagorean property with the Kullback-Leibler divergence
\cite{200206ISIT_Sun}. Moreover, the best guessing strategy is based
on a PMF that is a mixture of sources in the uncertainty set,
analogous to the source compression case. The min-sup value of
redundancy for the problem of compression under source uncertainty
is given by the capacity of a channel \cite{197909TR_Gal} with
inputs corresponding to the indices of the uncertainty set, and
channel transition probabilities given by the various sources in the
uncertainty set. We show that a similar result holds for guessing
under source uncertainty. In particular, the min-sup value is the
{\em channel capacity of order 1/$\alpha$} \cite{Csiszar-95} of an
appropriately defined channel.

The following is an outline of the paper. In Section
\ref{InaccuracyAndRedundancyInGuessing} we review known results for
the problem of guessing, introduce the relevant measures that
quantify redundancy, and show the relationship between this
redundancy and the divergence quantity $L_{\alpha}$. In Section
\ref{CampbellCodingTheorem}, we see how the same quantities arise in
the context of Campbell's minimum average exponential coding length
problem. In Section \ref{problemStatement}, we pose the min-sup
problem of quantifying the worst-case redundancy and identify
another inf-sup problem in terms $L_{\alpha}$. In Section
\ref{propertiesOfDivergence} we study the relations between
$L_{\alpha}$ and other known divergence measures. In Section
\ref{LalphaCenterRadius} we identify the so-called \emph{center} and
\emph{radius} of an uncertainty set. In Section \ref{examples}, we
specialize our results to two examples: the family of discrete
memoryless sources on finite alphabets, and the family of
finite-state arbitrarily varying sources. We establish results on
the asymptotic redundancies of these two uncertainty sets. We
further refine the redundancy upper bound for the family of binary
memoryless sources. In Section \ref{LalphaProjection} we conduct a
further study of $L_{\alpha}$ divergence and show that it satisfies
the Pythagorean property. Section \ref{conclusions} closes the paper
with some concluding remarks.

\section{Inaccuracy and redundancy in guessing}

\label{InaccuracyAndRedundancyInGuessing}

In this section, we prove previously known results in guessing. Our
aim is to motivate the study of quantities that measure inaccuracy
in guessing. In particular, we introduce a measure of divergence,
and show how it is related to the $\alpha$-divergence of Csisz\'{a}r
\cite{Csiszar-95}.

Let ${\mathbb{X}}$ and $\mathbb{Y}$ be finite alphabet sets.
Consider a correlated pair of random variables $(X,Y)$ with joint
PMF $P$ on $\mathbb{X} \times \mathbb{Y}$. Given side information $Y
= y$, we would like to guess the realization of $X$. Formally, a
guessing list $G$ with side information is a function $G: \mathbb{X}
\times \mathbb{Y} \rightarrow \{ 1, 2, \cdots, |\mathbb{X}| \}$ such
that for each $y \in \mathbb{Y}$, the function $G(\cdot,
y):\mathbb{X} \rightarrow \{ 1, 2, \cdots, |\mathbb{X}| \}$ is a
one-to-one function that denotes the order in which the elements of
$\mathbb{X}$ will be guessed when the guesser observes $Y = y$.
Naturally, knowing the PMF $P$, the best strategy which minimizes
the expected number of guesses, given $Y=y$, is to guess in the
decreasing order of $P(\cdot, y)$-probabilities. Let us denote such
an order $G_P$. Due to lack of exact knowledge of $P$, suppose we
guess in the decreasing order of probabilities of another PMF $Q$.
This situation leads to mismatch. In this section, we analyze the
performance of guessing strategies under mismatch.

In some of the results we will have $\rho > 0$, and in others $\rho
> -1, \rho \neq 0$. The $\rho >0$ case is of primary interest in the
context of guessing. The other case is also of interest in
Campbell's average exponential coding length problem where similar
quantities are involved.

Following the proof in \cite{Arikan}, we have the following simple
result for guessing under mismatch.

\vspace*{.1in}
\begin{proposition} ({\em Guessing under mismatch})
\label{guessingUpperBound} Let $\rho
>0$. Consider a source pair $(X,Y)$ with
PMF $P$. Let $Q$ be another PMF with $\mbox{Supp}(Q) = \mathbb{X}
\times \mathbb{Y}$. Let $G_Q$ be the guessing list with side
information $Y$ obtained under the assumption that the PMF is Q,
with ties broken using an arbitrary but fixed rule. Then the
guessing moment for the source with PMF $P$ under $G_Q$ satisfies
\begin{eqnarray}
\lefteqn{ \frac{1}{\rho} \log \left( \mathbb{E} \left[ G_Q(X,Y)^{\rho} \right] \right)} \nonumber \\
& \leq & \frac{1}{\rho} \log \left( \sum_{y \in \mathbb{Y}} \sum_{x
\in \mathbb{X}} P(x,y) \left[ \sum_{a \in {\mathbb{X}}} \left(
\frac{Q(a,y)}{Q(x,y)} \right)^{\frac{1}{1+\rho}} \right]^{\rho}
\right), \nonumber \\
\label{mism-eqn}
\end{eqnarray}
where the expectation $\mathbb{E}$ is with respect to $P$. $\hfill
\Box$
\end{proposition}
\vspace*{.1in}
\begin{proof}
For $\rho >0$, for each $y \in \mathbb{Y}$, observe that
\begin{eqnarray*}
G_Q(x,y) & \leq & \sum_{a \in {\mathbb{X}}} 1\{Q(a,y) \geq Q(x,y)\}
\\
& \leq & \sum_{a \in {\mathbb{X}}} \left( \frac{Q(a,y)}{Q(x,y)}
\right)^{\frac{1}{1+\rho}},
\end{eqnarray*}
for each $x \in \mathbb{X}$, which leads to the proposition.
\end{proof}
\vspace*{.1in}

For a source $P$ on $\mathbb{X} \times \mathbb{Y}$, the
\emph{conditional R\'{e}nyi entropy of order $\alpha$}, with $\alpha
> 0$, is given by
\begin{equation}
\label{conditionalRenyiEntropyFullFormula} H_{\alpha}(P) =
\frac{\alpha}{1-\alpha} \log \left( \sum_{y \in \mathbb{Y}} \left(
\sum_{x \in {\mathbb{X}}} P(x,y)^{\alpha} \right)^{1/\alpha}
\right).
\end{equation}
For the case when $|\mathbb{Y}| = 1$, {\it i.e.}, when there is no
side information, we may think of $P$ as simply a PMF on
$\mathbb{X}$. The above conditional R\'{e}nyi entropy of order
$\alpha$ is then the R\'{e}nyi entropy of order $\alpha$ of the
source $P$, given by
\begin{equation}
\label{RenyiEntropy} H_{\alpha}(P) = \frac{1}{1-\alpha} \log \left(
\sum_{x \in {\mathbb{X}}} P(x)^{\alpha} \right).
\end{equation}
Note that the left-hand side of
(\ref{conditionalRenyiEntropyFullFormula}) is written as a
functional of $P$ instead of the more common $H_{\alpha} \left( X
\mid Y \right)$. We do not use the latter because the dependence on
the PMF needs to be made explicit in many places in the sequel. Also
note that both (\ref{conditionalRenyiEntropyFullFormula}) and
(\ref{RenyiEntropy}) define $H_{\alpha}(P)$, one in the two random
variable case, and the other in the single random variable case. The
actual definition being referred to will be clear from the context.
It is well-known that
\begin{equation}
\label{conditionalRenyiEntropyBounds} 0 \leq H_{\alpha}(P) \leq \log
|\mathbb{X}|.
\end{equation}

Suppose that our guessing order is ``matched'' to the source, {\it
i.e.}, we guess according to the list $G_P$. We then get the
following corollary.

\vspace*{.1in}
\begin{corollary} ({\it Matched guessing, Arikan \cite{Arikan}})
\label{matchedGuessing} Under the hypotheses in Proposition
\ref{guessingUpperBound}, the guessing strategy $G_P$ satisfies
\begin{equation}
\label{matched-eqn} \frac{1}{\rho} \log \left( \mathbb{E} \left[
G_P(X,Y)^{\rho} \right] \right) \leq H_{\alpha}(P),
\end{equation}
where $\alpha = 1 / (1+\rho)$. $\hfill \Box$
\end{corollary}
\vspace*{.1in}

\begin{proof}
Set $Q=P$ in Proposition \ref{guessingUpperBound}.
\end{proof}
\vspace*{.1in}

Let us now look at the converse direction.

\vspace*{.1in}
\begin{proposition} ({\em Converse})
\label{guessingLowerBound}  Let $\rho > 0$. Consider a source pair
$(X,Y)$ with PMF $P$. Let $G$ be an arbitrary guessing list with
side information $Y$. Then, there is a PMF $Q_G$ on $\mathbb{X}
\times \mathbb{Y}$ with $\mbox{Supp}(Q_G) = \mathbb{X} \times
\mathbb{Y}$, and
\begin{eqnarray}
\lefteqn{ \frac{1}{\rho} \log \left( \mathbb{E} \left[
G(X,Y)^{\rho} \right] \right) } \nonumber \\
& \geq & \frac{1}{\rho} \log \left( \sum_{y \in \mathbb{Y}} \sum_{x
\in \mathbb{X}} P(x,y) \left[ \sum_{a \in {\mathbb{X}}} \left(
\frac{Q_G(a,y)}{Q_G(x,y)}
\right)^{\frac{1}{1+\rho}} \right]^{\rho} \right) \nonumber \\
\label{guessingLowerBoundEqn}&& - \log (1+ \ln |{\mathbb{X}}|),
\end{eqnarray}
where the expectation $\mathbb{E}$ is with respect to $P$. $\hfill
\Box$
\end{proposition}
\vspace*{.1in}
\begin{proof}
The proof is very similar to that of \cite[Theorem 1]{Arikan}.
Observe that because $\rho
>0$, for each $y \in \mathbb{Y}$, we have
\[
\sum_{x \in {\mathbb{X}}} \left( \frac{1}{G(x,y)} \right)^{1+\rho} =
\sum_{i=1}^{|{\mathbb{X}}|} \frac{1}{i^{1+\rho}} = c < \infty.
\]
Define the PMF $Q_G$ as
\[
Q_G(x,y) = \frac{1}{|\mathbb{Y}|} \cdot \frac{1}{c G(x,y)^{1+\rho}},
~~~~ \forall (x,y) \in \mathbb{X} \times \mathbb{Y}.
\]
Note that $\mbox{Supp}(Q_G) = \mathbb{X} \times \mathbb{Y}$.
Clearly, guessing in the decreasing order of $Q_G$-probabilities
leads to the guessing order $G$. By virtue of the definition of
$Q_G$, we have
\begin{eqnarray}
\lefteqn{ \sum_{y \in {\mathbb{Y}}} \sum_{x \in {\mathbb{X}}} P(x,y)
\left[ \sum_{a \in {\mathbb{X}}}
\left( \frac{Q_G(a,y)}{Q_G(x,y)} \right)^{\frac{1}{1+\rho}} \right]^{\rho}} \nonumber \\
& = & \sum_{y \in {\mathbb{Y}}} \sum_{x \in {\mathbb{X}}} P(x,y)
G(x,y)^\rho \cdot
\left( \sum_{a \in {\mathbb{X}}} \frac{1}{G(a,y)} \right)^{\rho} \nonumber \\
\label{converse} & \leq & \left( \sum_{y \in \mathbb{Y}} \sum_{x \in
{\mathbb{X}}} P(x,y) G(x,y)^\rho \right) \cdot \left( 1 + \ln
|\mathbb{X}| \right)^{\rho},
\end{eqnarray}
where the last inequality follows from (as in \cite{Arikan})
\[
\sum_{a \in {\mathbb{X}}} \frac{1}{G(a,y)} =
\sum_{i=1}^{|{\mathbb{X}}|} \frac{1}{i} \leq 1 + \ln |{\mathbb{X}}|,
~~ \forall y \in {\mathbb{Y}}.
\]
The proposition follows from (\ref{converse}).
\end{proof}
\vspace*{.1in}

Observe the similarity of the terms in the right-hand sides of
equations (\ref{mism-eqn}) and (\ref{guessingLowerBoundEqn}) in
Propositions \ref{guessingUpperBound} and \ref{guessingLowerBound},
respectively. The analog of this term in mismatched source
compression is $- \sum_{x \in \mathbb{X}} P(x) \log Q(x)$, which is
the expected length of a codebook built using a mismatched PMF $Q$.
The Shannon inequality (see, for example, \cite{1991EIT_CovTho})
states that
\[ - \sum_{x \in {\mathbb{X}}} P(x) \log Q(x) \geq -
\sum_{x \in {\mathbb{X}}} P(x) \log P(x) = H(P)
\]

The next inequality is analogous to the Shannon inequality. We can
interpret this as follows: if we guess according to some mismatched
distribution, then the expected number of guesses can only be
larger. We will let $\alpha = 1/(1+\rho)$ and expand the range of
$\alpha$ to $0 < \alpha < \infty$. A special case (when no side
information is available) was shown by Fischer (cf. \cite[Theorem
1.3]{Fis-Prag78}).

\vspace*{.1in}
\begin{proposition} ({\it Analog of Shannon inequality})
\label{ShannonInequalityAnalog}  Let $\alpha = \frac{1}{1+\rho} > 0,
\alpha \neq 1$. Then
\begin{eqnarray}
\label{ShannonInequalityAnalogEquation} \frac{\alpha}{1 - \alpha}
\log \left( \sum_{y \in {\mathbb{Y}}} \sum_{x \in {\mathbb{X}}}
P(x,y) \left[ \sum_{a \in {\mathbb{X}}} \left( \frac{Q(a,y)}{Q(x,y)}
\right)^{\alpha}
\right]^{\frac{1-\alpha}{\alpha}} \right) \nonumber \\
\geq H_{\alpha}(P),
\end{eqnarray}
with equality if and only if $P=Q$. $\hfill \Box$
\end{proposition}
\vspace*{.1in}
\begin{proof}
We will prove this directly using Holder's inequality. The right
side of (\ref{ShannonInequalityAnalogEquation}) is bounded. Without
loss of generality, we may assume that the left side of
(\ref{ShannonInequalityAnalogEquation}) is finite, for otherwise the
inequality trivially holds and $P \neq Q$. We may therefore assume
$\mbox{Supp}(P) \subset \mbox{Supp}(Q)$ under $0 < \alpha < 1$, and
$\mbox{Supp}(P) \cap \mbox{Supp}(Q) \neq \emptyset$ under $1 <
\alpha < \infty$ which are the conditions when the left side of
(\ref{ShannonInequalityAnalogEquation}) is finite.

With $\alpha = 1/(1+\rho)$, (\ref{ShannonInequalityAnalogEquation})
is equivalent to
\begin{eqnarray*}
\lefteqn{ \mbox{sign}(\rho) \cdot \sum_{y \in {\mathbb{Y}}} \sum_{x
\in {\mathbb{X}}} P(x,y) \left[ \sum_{a \in {\mathbb{X}}} \left(
\frac{Q(a,y)}{Q(x,y)}
\right)^{\frac{1}{1+\rho}} \right]^{\rho} } \\
& \geq & \mbox{sign}(\rho) \cdot \sum_{y \in \mathbb{Y}} \left(
\sum_{x \in \mathbb{X}} P(x,y)^{\frac{1}{1+\rho}} \right)^{1+\rho}.
\end{eqnarray*}
The above inequality holds term by term for each $y \in \mathbb{Y}$,
a fact that can be verified by using the H$\ddot{\mbox{o}}$lder
inequality
\begin{equation}
\label{HolderInequality} \mbox{sign}(\lambda) \cdot \left( \sum_x
u_x \right)^{\lambda} \cdot \left( \sum_x v_x \right)^{1 - \lambda}
\geq \mbox{sign} (\lambda) \cdot \left( \sum_x u_x^{\lambda} v_x^{1
- \lambda} \right)
\end{equation}
with $\lambda = \rho/(1+\rho) = 1 - \alpha$, $u_x =
Q(x,y)^{1/(1+\rho)}$,
\[
v_x = P(x,y)Q(x,y)^{-\rho/(1+\rho)},
\]
and raising the resulting inequality to the power $1 + \rho > 0$.
From the condition for equality in (\ref{HolderInequality}),
equality holds in (\ref{ShannonInequalityAnalogEquation}) if and
only if $P = Q$.
\end{proof}
\vspace*{.1in}

Proposition \ref{ShannonInequalityAnalog} motivates us to define the
following quantity that will be the focus of this paper:
\begin{eqnarray}
\lefteqn{ L_{\alpha}(P,Q) ~~ \stackrel{\Delta}{=} } \nonumber \\
& & \frac{\alpha}{1 - \alpha} \log \left( \sum_{y \in \mathbb{Y}}
\sum_{x \in {\mathbb{X}}} P(x,y) \left[ \sum_{a \in {\mathbb{X}}}
\left( \frac{Q(a,y)}{Q(x,y)} \right)^{\alpha}
\right]^{\frac{1-\alpha}{\alpha}} \right) \nonumber \\
\label{L-alpha} && ~~~~~~ - ~~ H_{\alpha}(P).
\end{eqnarray}
Proposition \ref{ShannonInequalityAnalog} indicates that
$L_{\alpha}(P,Q) \geq 0$, with equality if and only if $P = Q$.

Just as Shannon inequality can be employed to show the converse part
of the source coding theorem, we employ Proposition
\ref{ShannonInequalityAnalog} to get the converse part of a guessing
theorem. We thus have a slightly different proof of \cite[Theorem
1(a)]{Arikan}.

\vspace*{.1in}
\begin{thm} {\em (Arikan's Guessing Theorem \cite{Arikan}) }
\label{ArikanGuessingTheorem} Let $\rho > 0$. Consider a source pair
$(X,Y)$ with PMF $P$. Let $\alpha = \frac{1}{1+\rho}$. Then
\begin{eqnarray*}
\lefteqn{ H_{\alpha}(P) - \log (1+ \ln |{\mathbb{X}}|) }
\nonumber \\
& \leq & \frac{1}{\rho} \log \left( \min_{G} \mathbb{E} \left[
G(X,Y)^{\rho} \right] \right) \\
& \leq & H_{\alpha}(P).
\end{eqnarray*}
$\hfill \Box$
\end{thm}
\vspace*{.1in}

\begin{proof}
It is easy to see that the minimum is attained when the guessing
list is $G_P$, {\it i.e.}, when guessing proceeds in the decreasing
order of $P$-probabilities. Application of Proposition
\ref{guessingLowerBound} with $G = G_P$ and Proposition
\ref{ShannonInequalityAnalog} with $Q = Q_{G_P}$ yields the first
inequality. The upper bound follows from Corollary
\ref{matchedGuessing}.
\end{proof}
\vspace*{.1in}

{\it Remarks}: 1) $Q_{G_P}$ may be different from $P$ even though
they lead to the same guessing order.

2) Theorem \ref{ArikanGuessingTheorem} gives an operational meaning
to $H_{\alpha}(P)$; it indicates the exponent of the minimum
guessing moment to within $\log(1 + \ln|\mathbb{X}|)$.

3) Loosely speaking, Proposition \ref{ShannonInequalityAnalog}
indicates that mismatched guessing will perform worse than matched
guessing. The looseness is due to the looseness of the bound in
Theorem \ref{ArikanGuessingTheorem}.

Suppose now that we use an arbitrary guessing strategy $G$ to guess
$X$ with side information $Y$, when the source $(X,Y)$'s PMF is $P$.
$G$ may not necessarily be matched to the source, as would be the
case when the source statistics is unknown. Let us define its {\it
redundancy} in guessing $X$ with side information $Y$ when the
source is $P$ as follows:
\begin{equation}
\label{guessingRedundancyDefn}
  R(P,G) \stackrel{\Delta}{=} \frac{1}{\rho} \log \left( \mathbb{E} \left[ G(X,Y)^\rho \right]
  \right) - \frac{1}{\rho} \log \left( \mathbb{E} \left[ G_P(X,Y)^\rho \right]
  \right)
\end{equation}
The dependence of $R(P,G)$ on $\rho$ is understood and suppressed.
The following proposition bounds the redundancy on either side.

\vspace*{.1in}
\begin{thm}
\label{guessingRedundancyTheorem} Let $\rho > 0$, $\alpha =
1/(1+\rho)$. Consider a source pair $(X,Y)$ with PMF $P$. Let $G$ be
an arbitrary guessing list with side information $Y$ and $Q_G$ the
associated PMF given by Proposition \ref{guessingLowerBound}. Then
\begin{equation}
\label{guessingRedundancyTheoremEqn} \left| R(P,G) - L_{\alpha}(P,
Q_G) \right| \leq \log(1+\ln|\mathbb{X}|).
\end{equation}
$\hfill \Box$
\end{thm}
\vspace*{.1in}

\begin{proof}
The inequality $R(P,G) \leq L_{\alpha}(P,Q_G) +
\log(1+\ln|\mathbb{X}|)$ follows from Proposition
\ref{guessingUpperBound} applied with $Q = Q_G$, the first
inequality of Theorem \ref{ArikanGuessingTheorem}, and
(\ref{L-alpha}).

The inequality $R(P,G) \geq L_{\alpha}(P,Q_G) -
\log(1+\ln|\mathbb{X}|)$ follows from Proposition
\ref{guessingLowerBound}, the second inequality of Theorem
\ref{ArikanGuessingTheorem}, and (\ref{L-alpha}).
\end{proof}

\vspace*{.1in}

\emph{Remark}: It is possible that $P$ and $Q$ lead to the same
guessing order, \emph{i.e.}, $G_P = G_Q$. Thus $R(P,G_P) = R(P,G_Q)
= 0$. Yet, it is possible that $L_{\alpha}(P,Q)$ and
$L_{\alpha}(P,Q_{G_Q})$ are nonzero. This remains consistent with
Theorem \ref{guessingRedundancyTheorem} since
(\ref{guessingRedundancyTheoremEqn}) only provides bounds for
$R(P,G_Q)$ on either side to within $\log \left( 1 + \ln
|\mathbb{X}|\right)$, and is not an entirely accurate measure of
$R(P,G_Q)$. One can only conclude that
\[
L_{\alpha}(P,Q_{G_Q}) \leq \log (1 + \ln |\mathbb{X}|).
\]
In source compression with mismatch where the ``nuisance'' term is
not $\log \left( 1 + \ln |\mathbb{X}|\right)$ but the constant 1.
Yet, in the examples in Section \ref{examples} on guessing we see
how to make good use of these bounds. See also the discussion
following Theorem \ref{CampbellRedundancyTheorem} at the end of the
next section.

\section{Campbell's coding theorem and redundancy}

\label{CampbellCodingTheorem}

Campbell in \cite{Campbell-1} and \cite{Campbell-2} gave another
operational meaning to the R\'{e}nyi entropy of order $\alpha > 0$.
In this section we show that $L_{\alpha}$ arises as ``inaccuracy''
in this problem as well, when we encode according to a mismatched
source. To be consistent with the development in the previous
section, we will assume that $X$ is coded when the source coder has
side information $Y$.

Let $\mathbb{X}$ and $\mathbb{Y}$ be finite alphabet sets as before.
Let the true source probabilities be given by the PMF $P$ on
$\mathbb{X} \times \mathbb{Y}$. We wish to encode each realization
of $X$ using a variable-length code, given side information $Y$.
More precisely, let the (nonnegative) integer code lengths, $l(x,y)$
satisfy the Kraft inequality,
\[ \sum_{x \in \mathbb{X}} 2^{ -l(x,y) } \leq 1, ~~ \forall y \in \mathbb{Y} \]
The problem is then to choose $l$ among those that satisfy the Kraft
inequality so that the following is minimized:
\begin{equation}
\label{campbell-minimize} \frac{1}{\rho} \log \left( \mathbb{E}
\left[ 2^{ \rho l(X,Y) } \right] \right),~~ -1 < \rho < \infty, \rho
\neq 0,
\end{equation}
where the expectation $\mathbb{E}$ is with respect to the PMF $P$.
As $\rho \rightarrow 0$, this quantity tends to the expected length
of the code, $\mathbb{E}[l(X,Y)]$.

Observe that we may assume that $\sum_{x \in \mathbb{X}} 2^{-l(x,y)}
> 1/2$ for each $y$; otherwise we can reduce all lengths uniformly by 1, still
satisfy the Kraft inequality and get a strictly smaller value for
(\ref{campbell-minimize}). Henceforth, we focus only on length
functions that satisfy
\begin{equation}
\label{extendedKraftInequality} \frac{1}{2} < \sum_{x \in
\mathbb{X}} 2^{-l(x,y)} \leq 1, ~~ \forall y \in \mathbb{Y}.
\end{equation}

\vspace*{.1in}
\begin{thm} ({\it Campbell's Coding Theorem, Campbell \cite{Campbell-1}})
Let $-1 < \rho < \infty$, $\rho \neq 0$. Consider a source with PMF
$P$. Let $\alpha = \frac{1}{1+\rho}$. Then
\begin{eqnarray*}
H_{\alpha}(P)  \leq  \frac{1}{\rho} \log \left( \min_{l} \mathbb{E}
\left[ 2^{ \rho l(X,Y) } \right] \right) \leq H_{\alpha}(P) + 1,
\end{eqnarray*}
where the minimization is over all those length functions that
satisfy (\ref{extendedKraftInequality}). $\hfill \Box $
\end{thm}
\vspace*{.1in}

For a PMF $Q$ on $\mathbb{X} \times \mathbb{Y}$, let $l_Q$ be
defined by
\begin{eqnarray}
\label{mismatchedLengthFunction} l_Q(x,y) & \stackrel{\Delta}{=} &
\left\lceil -\log \left( \frac{Q(x,y)^{\frac{1}{1+\rho}}} {\sum_{a
\in \mathbb{X}} Q(a,y)^{\frac{1}{1+\rho}}} \right) \right\rceil \\
& = & \left\lceil - \log \left( Q'(x \mid y) \right) \right\rceil,
\end{eqnarray}
where $\lceil \cdot \rceil$ refers to the ceiling function and
$Q'(\cdot \mid y)$ is a conditional PMF on $\mathbb{X}$. Clearly,
$l_Q$ satisfies (\ref{extendedKraftInequality}).

Analogously, for any length function satisfying
(\ref{extendedKraftInequality}), we can define a PMF on $\mathbb{X}
\times \mathbb{Y}$ as follows:
\begin{equation}
\label{mismatchedPMF} Q_l(x,y) = \frac{1}{|\mathbb{Y}|} \frac{2^{-
(1+ \rho) l(x,y)}}{\sum_{a \in \mathbb{X}}2^{- (1+ \rho) l(a,y)} }.
\end{equation}
We can easily check that $l_{Q_l} = l$.

Let us define the redundancy for any $l$ satisfying
(\ref{extendedKraftInequality}) as
\begin{eqnarray*}
  \lefteqn{ R_c(P, l) } \nonumber \\
  & \stackrel{\Delta}{=} & \frac{1}{\rho} \log \left( \mathbb{E} \left[ 2^{\rho l(X,Y)} \right]
  \right) - \frac{1}{\rho} \log \left( \min_g \mathbb{E} \left[ 2^{\rho g(X,Y)} \right]
  \right),
\end{eqnarray*}
analogous to the definition without side information in
\cite{198809TIT_BluMcE}. Following the same sequence of steps as in
the mismatched guessing problem, it is straightforward to show the
following:

\vspace*{.1in}
\begin{thm}
\label{CampbellRedundancyTheorem} Let $-1 < \rho < \infty$, $\rho
\neq 0$, $\alpha = 1/(1+\rho)$. Consider a source pair $(X, Y)$ with
PMF $P$ on $\mathbb{X}$. Let $l$ be a length function that denotes
an encoding of $X$ with side information $Y$, and $Q_l$ the
associated PMF given by (\ref{mismatchedPMF}). Then
\begin{equation}
\label{CampbellRedundancyTheoremEqn} \left| R_c(P, l) -
L_{\alpha}(P, Q_l) \right| \leq 1.
\end{equation}
$\hfill \Box$
\end{thm}
\vspace*{.1in}

We interpret $L_{\alpha}(P, Q_l)$ as the penalty for mismatched
coding when $Q_l$ is not matched to $P$. $L_{\alpha}(P, Q_l)$ is
indicative of the redundancy to within a constant, as the
Kullback-Leibler divergence is in mismatched source compression. By
comparing (\ref{CampbellRedundancyTheoremEqn}) with
(\ref{guessingRedundancyTheoremEqn}), we see that the nuisance term
in this problem is a constant that does not depend on the size of
the source alphabet; $L_{\alpha}(P, Q_l)$ is therefore a more
faithful representation of $R_c(P,l)$ than $L_{\alpha}(P,Q_G)$ is of
$R(P,G)$.

\section{Problem statement}

\label{problemStatement}

Let $\mathbb{T}$ denote a set of PMFs on the finite alphabet
$\mathbb{X} \times \mathbb{Y}$. $\mathbb{T}$ may be infinite in
size. Associated with $\mathbb{T}$ is a family ${\cal{T}}$ of
measurable subsets of $\mathbb{T}$ and thus $(\mathbb{T},
{\cal{T}})$ is a measurable space. We assume that for every $(x,y)
\in \mathbb{X} \times \mathbb{Y}$, the mapping $P \mapsto P(x,y)$ is
$\cal{T}$-measurable.

For a fixed $\rho > 0$, we seek a good guessing strategy $G$ that
works well for all $P \in \mathbb{T}$. $G$ can depend on knowledge
of $\mathbb{T}$, but not on the actual source PMF. More precisely,
for $P \in \mathbb{T}$ the redundancy denoted by $R(P, G)$ when the
true source is $P$ and when the guessing list is $G$, is given by
(\ref{guessingRedundancyDefn}). The worst redundancy under this
guessing strategy is given by
\[
\sup_{P \in \mathbb{T}} R(P, G)
\]
Our aim is to minimize this worst redundancy over all guessing
strategies, {\it i.e.}, find a $G$ that attains the minimum
\begin{equation}
\label{minSupR} R^* = \min_{G} \sup_{P \in \mathbb{T}} R(P, G)
\end{equation}

In view of Theorem \ref{guessingRedundancyTheorem}, clearly, the
following quantity is relevant for $0 < \alpha < 1$. The definition
however is wider in scope.

\vspace*{.1in}
\begin{definition}
\label{minSupLalpha} For $0 < \alpha < \infty, \alpha \neq 1$,
\begin{equation}
\label{minSupLalphaEqn} C \stackrel{\Delta}{=} \min_Q \sup_{P \in
\mathbb{T}} L_{\alpha}(P, Q).
\end{equation}
\end{definition}
\vspace*{.1in} The following theorem justifies the use of ``min''
instead of ``inf''.

\vspace*{.1in}
\begin{thm}
\label{minSupLalphaTheorem} There exists a unique PMF $Q^*$ such
that
\[
C = \sup_{P \in \mathbb{T}} L_{\alpha}(P, Q^*) = \inf_Q \sup_{P \in
\mathbb{T}} L_{\alpha}(P, Q).
\]
$\hfill \Box$
\end{thm}
\vspace*{.1in}

The proof is in Section \ref{ProofOfExistenceTheorem}.

\vspace*{.1in} {\em Remark}: 1) $C \leq \log |\mathbb{X}|$ and is
therefore finite. Indeed, take $Q$ to be uniform PMF on $\mathbb{X}
\times \mathbb{Y}$. Then
\[
L_{\alpha}(P, Q) = \log |\mathbb{X}| - H_{\alpha}(P) \leq \log
|\mathbb{X}|, ~ \forall P \in {\mathbb{T}}.
\]

2) The minimizing $Q^*$ has the geometric interpretation of a {\it
center} of the uncertainty set $\mathbb{T}$. Accordingly, $C$ plays
the role of {\it radius}; all elements in the uncertainty set
$\mathbb{T}$ are within a ``squared distance'' $C$ from the center
$Q^*$. The reason for describing $L_{\alpha}(P,Q)$ as ``squared
distance'' will become clear after Proposition
\ref{pythagoreanProposition}.

The following result shows how to find good guessing schemes under
uncertainty.
\vspace*{.1in}
\begin{thm} ({\em Guessing under uncertainty})
\label{guessingUnderUncertainty} Let $\mathbb{T}$ be a set of PMFs.
There exists a guessing list $G^*$ for $X$ with side information $Y$
such that
\[
\sup_{P \in \mathbb{T}} R(P, G^*) \leq C + \log(1 + \ln
|\mathbb{X}|).
\]
Conversely, for any arbitrary guessing strategy $G$, the worst-case
redundancy is at least $C - \log(1 + \ln |\mathbb{X}|)$, {\it i.e.},
\[
\sup_{P \in \mathbb{T}} R(P, G) \geq C - \log(1 + \ln |\mathbb{X}|).
\]
$\hfill \Box$
\end{thm}
\vspace*{.1in}

\begin{proof}
Let $Q^*$ be the PMF on $\mathbb{X} \times \mathbb{Y}$ that attains
the minimum in (\ref{minSupLalphaEqn}), {\it i.e.},
\begin{equation}
\label{minMaxValue} C = \sup_{P \in \mathbb{T}}L_{\alpha}(P,Q^*).
\end{equation}

Let $G^* = G_{Q^*}$. Then
\begin{equation}
\label{perSourceRedundancy} R(P, G^*) \leq L_{\alpha}(P,Q^*) +
\log(1+\ln|\mathbb{X}|)
\end{equation}
follows from Proposition \ref{guessingUpperBound} applied with $Q =
Q^*$, the first inequality of Theorem \ref{ArikanGuessingTheorem},
and (\ref{L-alpha}), as in the proof of Theorem
\ref{guessingRedundancyTheorem}. After taking supremum over all $P
\in \mathbb{T}$, and after substitution of (\ref{minMaxValue}), we
get
\begin{eqnarray*}
\sup_{P \in \mathbb{T}} R(P, G^*) & \leq & \sup_{P \in
\mathbb{T}}L_{\alpha}(P,Q^*) +
\log(1+\ln|\mathbb{X}|) \\
& = & C + \log(1+\ln|\mathbb{X}|),
\end{eqnarray*}
which proves the first statement.

For any guessing strategy $G$, observe that Theorem
\ref{guessingRedundancyTheorem} implies that
\[
R(P, G) \geq L_{\alpha}(P,Q_G) - \log(1+\ln|\mathbb{X}|),
\]
and therefore
\begin{eqnarray*}
\sup_{P \in \mathbb{T}} R(P, G) & \geq & \sup_{P \in \mathbb{T}}
L_{\alpha}(P,Q_G) -
\log(1+\ln|\mathbb{X}|) \\
& \geq & C - \log(1+\ln|\mathbb{X}|),
\end{eqnarray*}
which proves the second statement.
\end{proof}
\vspace*{.1in}

{\em Remarks}: 1) Thus one approach to obtain the minimum in
(\ref{minSupR}) is to identify minimum value in
(\ref{minSupLalphaEqn}). This minimum value will be within
$\log(1+\ln|\mathbb{X}|)$ of $R^*$ in (\ref{minSupR}). Moreover, the
corresponding minimizer $Q^*$ can be used to generate a guessing
strategy.

2) Theorem \ref{guessingUnderUncertainty} can be easily restated for
Campbell's coding problem. The nuisance term
$\log(1+\ln|\mathbb{X}|)$ is now replaced by the constant 1.

3) The converse part of Theorem \ref{guessingUnderUncertainty} is
meaningful only when $C > \log(1 + \ln |\mathbb{X}|)$. This will
hold, for example, when the uncertainty set is sufficiently rich.
The finite state, arbitrarily varying source is one such example.
Observe that if we have $\mathbb{X} \times \mathbb{Y} = \mathbb{A}^n
\times \mathbb{B}^n$, then $\log(1 + \ln |\mathbb{X}|)$ grows
logarithmically with $n$ if $|\mathbb{X}| \geq 2$. The uncertainty
set will be rich enough for the converse to be meaningful if $C$
grows with $n$ at a faster rate.

\section{Relations between $L_{\alpha}$ and other divergence quantities}

\label{propertiesOfDivergence}

Having shown how $L_{\alpha}(P,Q)$ arises as a penalty function for
mismatched guessing and coding, we now study it in greater detail
and relate it to other divergence quantities. The relationships we
discover here will be useful in the sequel. Throughout this section,
$0 < \alpha < \infty, \alpha \neq 1$. Accordingly, $-1 < \rho <
\infty, \rho \neq 0$. Let $P$ and $Q$ be PMFs on $\mathbb{X} \times
\mathbb{Y}$.

\begin{enumerate}
\item As we saw before, $L_{\alpha}(P,Q) \geq 0$, with equality if and only if $P=Q$.

\item $L_{\alpha}(P,Q) = \infty$ if and only if $\mbox{Supp}(P) \cap
\mbox{Supp}(Q) = \emptyset$, or $\alpha < 1$ and $\mbox{Supp}(P)
\not\subset \mbox{Supp}(Q)$.

\item Given the joint PMF $P$, let us define the ``tilted''
conditional PMF on $\mathbb{X}$ as follows:
\begin{equation}
\label{tilted} P'(x \mid y) \stackrel{\Delta}{=}
 \left\{
 \begin{array}{ll}
 P(x,y)^{\alpha} ~/~ \sum_{a \in {\mathbb{X}}}P(a,y)^{\alpha},\\
 \\
 ~~~~~~~~ \mbox{ if } \sum_{a \in {\mathbb{X}}}P(a,y)^{\alpha} > 0, \\
 \\
 1/|\mathbb{X}|, \mbox{ otherwise.}
 \end{array}
 \right.
\end{equation}
The above definition simplifies many expressions in the sequel. The
dependence on $\alpha$ in the mapping $P \mapsto P'$ is suppressed.

\item When $|\mathbb{Y}| = 1$, we interpret that no side
information is available. Then $P$ and $Q$ may be thought of PMFs on
$\mathbb{X}$ with no reference to $\mathbb{Y}$. $P'$ and $Q'$ given
by (\ref{tilted}) are PMFs in one-to-one correspondence with $P$ and
$Q$ respectively.

Using the expression for R\'{e}nyi entropy and (\ref{L-alpha}), we
have that
\begin{eqnarray}
L_{\alpha}(P,Q) & = & \frac{1}{\rho} \log \left( \sum_{x \in
{\mathbb{X}}} P'(x)^{1+\rho} \cdot Q'(x)^{-\rho}
\right) \nonumber \\
\label{LDrelation}& = & D_{1/\alpha}(P' \parallel Q'),
\end{eqnarray}
where $D_{\beta}(R \parallel S)$ is the R\'{e}nyi's information
divergence of order $\beta$,
\[ D_{\beta}(R \parallel S) = \frac{1}{\beta -1} \log \left( \sum_{x \in
{\mathbb{X}}} R(x)^{\beta} S(x)^{1-\beta} \right), \] which is $\geq
0$ and equals 0 if and only if $R=S$. For the case when
$|\mathbb{Y}| = 1$ we therefore have another proof of Proposition
\ref{ShannonInequalityAnalog}.

\item The conditional Kullback-Leibler divergence is recovered as follows:
\begin{eqnarray*}
\lim_{\alpha \rightarrow 1} L_{\alpha}(P,Q) & = & \sum_{y} \sum_{x}
P(x,y) \log \left( \frac{P(x \mid y)}{Q(x \mid y)} \right),
\end{eqnarray*}
where $Q(\cdot \mid y)$ and $P(\cdot \mid y )$ are the respective
conditional PMFs of $X$ given $Y = y$.

\item \label{comment:NotAConvexFunction} In general, $L_{\alpha}(P,Q)$ is
not a convex function of $P$. Moreover, it is not, in general, a
convex function of $Q$.

\item \label{comment:NoDataProcessingInequality} In general, $L_{\alpha}(P,Q)$ does not satisfy
the so-called data-processing inequality. More precisely, if
$\mathbb{X}'$ and $\mathbb{Y}'$ are finite sets, and if $f:
\mathbb{X} \times \mathbb{Y} \rightarrow \mathbb{X}' \times
\mathbb{Y}'$ is a function, it is not necessarily true that
$L_{\alpha}(P,Q) \geq L_{\alpha}(Pf^{-1},Qf^{-1})$.

\item When $|\mathbb{Y}| = 1$, {\it i.e.}, in the no side information case,
using (\ref{tilted}) we can write $L_{\alpha}(P,Q)$ as follows:
\begin{equation}
\label{LIrelation} L_{\alpha}(P,Q) = \frac{1}{\rho} \log \left[
\mbox{sign}(\rho) \cdot I_f(P' \parallel Q') \right],
\end{equation}
where $I_f(R \parallel S)$ is the $f$-divergence
\cite{Csiszar-Informativity} given by
\begin{equation}
\label{f-divergence} I_{f}(R \parallel S) = \sum_{x \in
{\mathbb{X}}} S(x) f\left( \frac{R(x)}{S(x)}\right),
\end{equation}
with
\begin{equation}
\label{f-definition} f(x) = \mbox{sign}(\rho) \cdot x^{1+\rho}, ~~ x
\geq 0.
\end{equation}
Since $f$ is a strictly convex function for $\rho \neq 0$, an
application of Jensen's inequality in (\ref{f-divergence}) indicates
that
\begin{equation}
\label{IfLowerBound} I_f(R  \parallel  S) \geq f(1) = \left\{
\begin{array}{rc} -1, & -1 < \rho < 0, \\ 1, & 0 < \rho < \infty.
\end{array} \right.
\end{equation}
Moreover, when $-1 < \rho < 0$, we have the following bounds:
\begin{equation}
\label{IfHighAlphaBounds} -1 \leq I_f(R  \parallel  S) \leq 0.
\end{equation}

\item Let us define
\[
h(P) \stackrel{\Delta}{=} \sum_{y \in \mathbb{Y}} \left( \sum_{x \in
\mathbb{X}} P(x,y)^{\alpha} \right)^{\frac{1}{\alpha}}.
\]
The dependence of $h$ on $\alpha$ is understood, and suppressed for
convenience. Clearly,
\begin{equation}
\label{conditionalRenyiEntropy} H_{\alpha}(P) = \frac{\alpha}{1 -
\alpha} \log h(P).
\end{equation}

Motivated by the relationship in (\ref{LIrelation}), let us write
$L_{\alpha}$ in the general case as follows:
\begin{equation}
\label{LIrelationGeneral} L_{\alpha}(P,Q) =  \frac{1}{\rho} \log
\left[ \mbox{sign}(\rho) \cdot I(P,Q) \right],
\end{equation}
where $I(P,Q)$ is given by
\begin{eqnarray}
\lefteqn{I(P,Q)} \nonumber \\
\label{IDefinitionRho} & \stackrel{\Delta}{=} &
\frac{\mbox{sign}(\rho)}{h(P)} \sum_{y \in \mathbb{Y}} \sum_{x \in
{\mathbb{X}}} P(x,y) \left( Q'(x
\mid y) \right)^{-\rho}, \\
& = & \frac{\mbox{sign}(1 - \alpha)}{h(P)} \sum_{y \in \mathbb{Y}}
\sum_{x \in {\mathbb{X}}} P(x,y) \left( Q'(x \mid y)
\right)^{\frac{\alpha - 1}{\alpha}}. \nonumber \\
\label{IDefinitionAlpha}
\end{eqnarray}
These expressions turn out to be useful in the sequel.

It is not difficult to show that
\[
I(P,Q) = \sum_{y \in \mathbb{Y}} w(y) \cdot I_f(P'(\cdot \mid y)
\parallel Q'(\cdot \mid y)),
\]
where $w$ is the PMF on $\mathbb{Y}$ given by
\[
w(y) = \frac{1}{h(P)} \cdot \left( \sum_{x \in \mathbb{X}}
P(x,y)^{\alpha} \right)^{\frac{1}{\alpha}}.
\]
Consequently, the bounds given in (\ref{IfLowerBound}) and
(\ref{IfHighAlphaBounds}) are valid for $I(P,Q)$, under
corresponding conditions on $\alpha$.

\item Inequalities involving $L_{\alpha}$ result in inequalities involving
$I$ with ordering preserved. More precisely, for $r \geq 0$, if
$L_{\alpha}(P,Q) < r$, then $I(P,Q) < t$, for $t = \mbox{sign}(\rho)
\cdot 2^{\rho r}$.

\item From the known bounds $0 \leq H_{\alpha}(P) \leq
\log |\mathbb{X}|$, it is easy to see the following bounds:
\begin{equation}
\label{scaledRenyiEntropy1} 1 \leq h(P) \leq
|\mathbb{X}|^{\frac{1-\alpha}{\alpha}}, \mbox{ for } 0 < \alpha < 1,
\end{equation}
and
\begin{equation}
\label{scaledRenyiEntropy2} |\mathbb{X}|^{\frac{1-\alpha}{\alpha}}
\leq h(P) \leq 1, \mbox{ for } 1 < \alpha < \infty.
\end{equation}
In both cases, we see that $h(P)$ is bounded away from 0 and
therefore (\ref{IDefinitionRho}) and (\ref{IDefinitionAlpha}) are
well-defined.
\end{enumerate}

The quantity $L_{\alpha}(P,Q)$ does not have many of the useful
properties enjoyed by the Kullback-Leibler divergence, or other
$f$-divergences, even in the case when $|\mathbb{Y}| = 1$. See for
example, comments \ref{comment:NotAConvexFunction} and
\ref{comment:NoDataProcessingInequality} made earlier in this
section. However, it behaves like squared distance and shares a
``Pythagorean'' property with the Kullback-Leibler divergence. This
is explored in Section \ref{LalphaProjection}.

\section{$L_{\alpha}$-center and radius of a family}

\label{LalphaCenterRadius}

In this section we identify the $L_{\alpha}$-center and radius of a
family. We first begin with a finite family and subsequently study
an arbitrary family (that satisfies some measurability conditions).
We finally conclude the section with a proof of Theorem
\ref{minSupLalphaTheorem}.

\subsection{$L_{\alpha}$-center and radius of a finite family}

\label{LalphaCenterRadiusFiniteFamily}

Let $|\mathbb{T}|$ be finite. For simplicity, assume that no side
information is available. We will therefore use $\mathbb{X}$ instead
of the cumbersome $\mathbb{X} \times \mathbb{Y}$. Our main goals
here are to verify using known results that the $L_{\alpha}$-center
exists, is unique, and lies is in the closure of the convex hull of
$\mathbb{T}$. We then briefly touch upon connections with Gallager
exponents, capacity of order $1/\alpha$, and information radius of
order $1/\alpha$. The development in this section will suggest an
approach to prove Theorem \ref{minSupLalphaTheorem} for the case
when $|\mathbb{T}|$ is infinite.

\subsubsection{Proof of Theorem \ref{minSupLalphaTheorem} for a finite family of PMFs}
Let $\mathbb{T} = \{ P_1, \cdots, P_m\}$ be PMFs on $\mathbb{X}$.
The problem of identifying the $L_{\alpha}$-center and radius can be
solved by identifying the $D_{1/\alpha}$-center and radius of the
tilted family of PMFs $\{ P'_i \mid 1 \leq i \leq m\}$, where the
invertible transformation from $Q \mapsto Q'$ is given by
(\ref{tilted}). Moreover, from (\ref{LDrelation}) and
(\ref{LIrelation}), we have
\begin{eqnarray}
\label{minMaxLalpha} \lefteqn{ \inf_{Q} \max_{1 \leq i \leq m}
L_{\alpha}(P_i, Q) } \\
\label{D-beta-minMax} & = & \inf_{Q} \max_{1 \leq i \leq m}
D_{1+\rho} (P'_i  \parallel  Q') \\
\label{I-f} & = & \frac{1}{\rho} \log \left( \mbox{sign}(\rho)~
\inf_{Q} \max_{1 \leq i \leq m} I_{f} (P'_i \parallel Q') \right),
\end{eqnarray}

Csisz\'{a}r considered the evaluation of (\ref{D-beta-minMax}) in
\cite[Proposition 1]{Csiszar-95}, and the evaluation of the inf-max
within parenthesis in (\ref{I-f}) in \cite{Csiszar-Informativity}.

From \cite[Theorem 3.2]{Csiszar-Informativity} and its Corollary
(the required conditions for their application are $f$ is strictly
convex and $f(0) < \infty$; these clearly hold since $\rho \neq 0$
and $f(0)=0$) there exists a unique PMF $(Q')^{*}$ on
${\mathbb{X}}$, which minimizes $\max_{1 \leq i \leq m} I_{f} (P'_i
\parallel Q')$. From the bijectivity of the $Q \mapsto Q'$ mapping, the
infima in (\ref{minMaxLalpha}), (\ref{D-beta-minMax}), and
(\ref{I-f}) can all be replaced by minima. From the inverse of the
map $Q \mapsto Q'$, we obtain the unique minimizer $Q^*$ for
(\ref{minMaxLalpha}). This proves the existence and uniqueness
result of Theorem \ref{minSupLalphaTheorem} when $|\mathbb{T}|$ is
finite.

\vspace*{.1in}

\subsubsection{Minimizer is in the convex hull}

\label{subsubsec:minimizerInConvexHull}

Let $\cal{E}$ be the convex hull of $\mathbb{T}$. That the minimizer
$Q^*$ is in the convex hull of the family, {\it i.e.}, $Q^* \in
\cal{E}$, can be gleaned from the results of \cite[Equation
2.25]{Csiszar-Informativity}, \cite[Theorem
3.2]{Csiszar-Informativity}, and its Corollary. Indeed,
\cite[Theorem 3.2]{Csiszar-Informativity} assures that
\begin{eqnarray}
\label{I-f1} \lefteqn{ \min_{Q'} \max_{1 \leq i \leq m} I_{f} (
P'_i \parallel Q') } \\
\label{maxMinObjective} & = & \max_{\mu} \min_{Q'} \sum_{i=1}^m
\mu(i) I_{f} (P'_i \parallel Q'),
\end{eqnarray}
where the max-min in (\ref{maxMinObjective}) is achieved at
$(\mu^{*}, Q'^{*})$, and $Q'^{*}$ is the PMF which attains the
min-max in (\ref{I-f1}). We now seek to find out the nature of
$Q'^{*}$ and thence $Q^*$.

For any arbitrary weight function $\mu$, we have from \cite[Equation
2.25]{Csiszar-Informativity} that the $Q'$ which minimizes
\begin{equation}
\label{averageIfObjective} \sum_{i=1}^m \mu(i) I_{f} (P'_i \parallel
Q')
\end{equation}
is
\begin{eqnarray}
\label{minimizer-informativity-primitive}Q'(x) & = & c^{-1} \cdot
\left(
\sum_{i=1}^m \mu(i) (P'_i(x))^{1/\alpha} \right)^{\alpha} \\
\label{minimizer-informativity} & = & c^{-1} \left( \sum_{i=1}^m
\frac{\mu(i)}{h(P_i)} P_i(x) \right)^{1/\alpha}
\end{eqnarray}
for every $x \in {\mathbb{X}}$, where $c$ is the normalizing
constant. From the correspondence between the primed and the
unprimed PMFs, and (\ref{minimizer-informativity}), we obtain
\begin{equation}
\label{modify-w} Q(x) = d^{-1} \sum_{i=1}^m \frac{\mu(i)}{h(P_i)}
P_i(x), ~~ \forall x \in {\mathbb{X}}
\end{equation}
where $d$ is the normalizing constant
\begin{equation}
\label{normalizer} d = \sum_{i=1}^m \frac{\mu(i)}{h(P_i)}.
\end{equation}
Thus, for an arbitrary $\mu$, the $Q$ (obtained from $Q'$) that
minimizes (\ref{averageIfObjective}) is in the convex hull
${\cal{E}}$. In particular, the minimizing $Q^*$ corresponding to
the $\mu^{*}$ that attains the max-min objective in
(\ref{maxMinObjective}), and therefore the min-max objective in
(\ref{I-f1}), is also in ${\cal{E}}$. This result will be proved in
wider generality in Section \ref{LalphaProjection}.

With some algebra, we can further show that
\begin{equation}
\label{minMaxLalphaExpression} C = \min_{Q} \max_{1 \leq i \leq m}
L_{\alpha} (P_i, Q) = \frac{\alpha}{1-\alpha} \log (d \cdot h(Q^*)),
\end{equation}
where $Q^*$ is given by (\ref{modify-w}) and $d$ by
(\ref{normalizer}) with $\mu = \mu^*$.

\vspace*{.1in}

\subsubsection{Necessary and sufficient conditions for finding the $L_{\alpha}$-center and radius}
\label{FiniteCaseNASC}

From \cite[Theorem 3.2]{Csiszar-Informativity}, a weight vector
$\mu$ maximizes (\ref{maxMinObjective}) if and only if
\begin{equation}
\label{NASC} I_{f}(P'_i \parallel Q') \leq K, ~~ i = 1, 2, \cdots,
m,
\end{equation}
where equality holds whenever $\mu(i) > 0$, and $Q'$ is given by
(\ref{minimizer-informativity-primitive}). Under this condition,
clearly, the corresponding $Q$ given by (\ref{modify-w}) is the
$L_{\alpha}$-center and $C = (1/\rho) \log(\mbox{sign}(\rho) \cdot
K)$ is the $L_{\alpha}$-radius.

An interesting special case occurs when $h(P_i)$ is independent of
$i$. Then we may simplify (\ref{modify-w}) to
\begin{equation}
\label{sameWtEquation} Q = \sum_{i=1}^m \mu(i) ~P_i,
\end{equation}
{\it i.e.}, the weights that make the optimum mixture (of PMFs) are
the same as the given weights that form the objective function in
(\ref{maxMinObjective}).

\vspace*{.1in}

\subsubsection{Relationship with Gallager exponent}

For the set of PMFs $\{ P_i \mid 1 \leq i \leq m \}$ the tilted set
$\{ P'_i \mid 1 \leq i \leq  m \}$ can be considered as a channel
with input alphabet $\{1, 2, \cdots , m \}$ and output alphabet
${\mathbb{X}}$. This channel will be represented as $P'$.

From the remarks in \cite{Csiszar-95} on the connection between
information radius of order $1/\alpha$ and the Gallager exponent of
the channel $P'$, and from \cite[Proposition 1]{Csiszar-95}, we have
\begin{eqnarray*}
\min_{Q} \max_{1 \leq i \leq m} L_{\alpha} (P_i, Q) = \max_{\mu}
\frac{1}{\alpha-1} ~ E_{o}(\alpha-1, \mu, P'),
\end{eqnarray*}
where the right-hand side is the maximized Gallager exponent of the
channel $P'$. ($1 < \alpha < 2$ is relevant in \cite[p.
138]{Gal-ITRC68}, $1 < \alpha < \infty $ in \cite[p.
157]{Gal-ITRC68}, and $0 < \alpha < 1$ in \cite{Arimoto}).

\vspace*{.1in}

\subsubsection{The max-min problem for $L_{\alpha}$}

Thus far our focus has been on the min-max problem of finding the
$L_{\alpha}$-center. We briefly looked at identifying the max-min
value of $I_f$ in (\ref{maxMinObjective}), but only as a means to
study the min-max problem. We now make some remarks about the
max-min problem for the finite family case. Its extension to
arbitrary uncertainty sets is not considered in this paper.

Suppose that our new objective is to find
\begin{equation}
\label{maxMinLalphaObjective} \max_{\mu} \min_{Q} \sum_{i=1}^m
\mu(i) L_{\alpha}(P_i, Q).
\end{equation}
This problem is the same as identifying the ``capacity of order
$1/\alpha$'' of the channel $P'$ \cite{Csiszar-95}, {\it i.e.},
\[
\max_{\mu} \min_{Q'} \sum_{i=1}^m \mu(i) D_{1/\alpha}(P'_i
\parallel  Q').
\]
\cite[Proposition 1]{Csiszar-95} solves this problem; the value is
the same as the min-max value
\[
  \min_{Q'} \max_{1 \leq i \leq m} D_{1/\alpha}(P'_i  \parallel  Q').
\]
Consequently, the max-min value of (\ref{maxMinLalphaObjective}) is
the same as the $L_{\alpha}$-radius of the family.

\subsection{$L_{\alpha}$-center and radius for an arbitrary family}

We are now back to the case with side information and an infinite
family $\mathbb{T}$. The development in this subsection will be
analogous to Gallager's approach \cite{197909TR_Gal} for source
compression. We first recall the technical condition indicated in
Section \ref{problemStatement}. $\mathbb{T}$ is a family of PMFs on
$\mathbb{X} \times \mathbb{Y}$, $(\mathbb{T}, {\cal{T}})$ a
measurable space, and for every $(x,y) \in \mathbb{X} \times
\mathbb{Y}$, the mapping $P \mapsto P(x,y)$ is $\cal{T}$-measurable.

Our focus will be on the following:
\vspace*{.1in}
\begin{definition}
\label{minSupIf} For $0 < \alpha < \infty, \alpha \neq 1$,
\begin{equation}
\label{minSupIfEqn} K_+ \stackrel{\Delta}{=} \min_Q \sup_{P \in
\mathbb{T}} I(P, Q).
\end{equation}
\end{definition}
\vspace*{.1in}

Taking $Q$ to be the uniform PMF on $\mathbb{X} \times \mathbb{Y}$
it is easy to check that $K_+$ is finite; indeed $1 \leq K_+ \leq
|\mathbb{X}|^{\rho}$ when $\rho
> 0$ and $-1 \leq K_+ \leq 0$ when $-1 < \rho < 0$.

Let us define some other auxiliary quantities. Define the mapping
$f:\mathbb{T} \rightarrow \mathbb{R}_+^{|\mathbb{X}| |\mathbb{Y}|}$
as follows:
\[
f(P) \stackrel{\Delta}{=} P / h(P).
\]
For a probability measure $\mu$ on $(\mathbb{T}, {\cal{T}})$, let
\begin{equation}
\label{outputPMFUnnormalized} F \stackrel{\Delta}{=}
\int_{\mathbb{T}} d \mu(P) f(P).
\end{equation}
We define the PMF $\mu f \in {\cal{P}}(\mathbb{X} \times
\mathbb{Y})$ as the scaled version of $F$,
\begin{eqnarray}
\label{outputPMF} \mu f \stackrel{\Delta}{=} d^{-1} F
\end{eqnarray}
where $d$ as in the finite case is the normalizing constant
\begin{equation}
\label{outputPMFNormalizer} d \stackrel{\Delta}{=} \int_{\mathbb{T}}
~ \frac{d\mu(P)}{h(P)} = \sum_{x \in \mathbb{X}} F(x).
\end{equation}
These definitions are extensions of (\ref{modify-w}) and
(\ref{normalizer}) to arbitrary $\mathbb{T}$. Moreover, let
\begin{equation}
\label{MI} J(\mu, \mathbb{T}) \stackrel{\Delta}{=} \int_{\mathbb{T}}
~ d\mu(P)~ I(P, \mu f).
\end{equation}
Simple algebraic manipulations result in
\begin{eqnarray}
\label{J-Fexpansion} J(\mu, \mathbb{T}) & = & \mbox{sign}(\rho) \cdot  h(F) \\
\label{J-simplification} & = & \mbox{sign}(\rho) \cdot d \cdot h(\mu
f),
\end{eqnarray}
an extension of \cite[Equation (2.24)]{Csiszar-Informativity} for
arbitrary $\mathbb{T}$.

The following auxiliary quantity will be useful. \vspace*{.1in}
\begin{definition}
\label{supMinIf} For $0 < \alpha < \infty, \alpha \neq 1$,
\begin{equation}
\label{supJEqn} K_- \stackrel{\Delta}{=} \sup_{\mu} J(\mu,
\mathbb{T}).
\end{equation}
\end{definition}
\vspace*{.1in}

The quantity $\mu f$ in (\ref{outputPMF}) is analogous to the PMF at
the output of a channel represented by $\mathbb{T}$ when the input
measure is $\mu$. $J(\mu, \mathbb{T})$ in (\ref{MI}) is the analogue
of mutual information; Csisz\'{a}r calls it informativity in his
work on finite-sized families \cite{Csiszar-Informativity}.

\vspace*{.1in}
\begin{proposition}
\label{K-K+relation} $K_- \leq K_+$.
\end{proposition}
\vspace*{.1in}
\begin{proof}
Fix an arbitrary PMF $Q$ on $\mathbb{X} \times \mathbb{Y}$. It is
straightforward to show that \cite[Equation
2.26]{Csiszar-Informativity} holds even when $|\mathbb{T}|$ is not
finite, and is given by
\[
\int_{\mathbb{T}} d\mu(P) ~ I(P,Q) = \mbox{sign}(\rho) \cdot J(\mu,
\mathbb{T}) \cdot I(\mu f, Q).
\]
Since $I(\mu f, Q) \geq \mbox{sign}(\rho)$, it follows that
\[
\int_{\mathbb{T}} d\mu(P) \cdot I(P, Q) \geq J(\mu, \mathbb{T}).
\]
Consequently
\[
J(\mu, \mathbb{T}) = \min_Q \int_{\mathbb{T}} d\mu(P) ~ I(P,Q),
\]
which leads to
\begin{eqnarray*}
K_- & = & \sup_{\mu} J(\mu, \mathbb{T}) \\
& = & \sup_{\mu} \min_Q \int_{\mathbb{T}} d\mu(P) ~
I(P,Q) \\
& \leq & \min_Q \sup_{\mu} \int_{\mathbb{T}} d\mu(P) ~
I(P,Q) \\
& = & \min_Q \sup_{P \in \mathbb{T}} I(P,Q) \\
& = & K^+.
\end{eqnarray*}
\end{proof}
\vspace*{.1in}

The following Proposition is similar to \cite[Theorem
A]{197909TR_Gal}. The proof largely runs along similar lines.

\vspace*{.1in}
\begin{proposition}
\label{GallagerExtn} A real number $R$ equals $K_-$ if and only if
there exist a sequence of probability measures $(\mu_n : n \in
\mathbb{N})$ on $(\mathbb{T}, \cal{T})$ and a PMF $Q^*$ on
$\mathbb{X} \times \mathbb{Y}$ with the following properties:
\begin{enumerate}
\item $ \lim_n J(\mu_n, \mathbb{T}) = R$;
\item $ \lim_n \mu_n f = Q^*$;
\item $ I(P, Q^*) \leq R$, for every $P \in
\mathbb{T}$.
\end{enumerate}
Furthermore $Q^*$ is unique, attains the minimum in
(\ref{minSupIfEqn}), and $K_- = K_+$. $\hfill \Box$
\end{proposition}
\vspace*{.1in}
\begin{proof}
$\Leftarrow$: Observe that on account of 1), 3), and Proposition
\ref{K-K+relation} we have
\begin{eqnarray*}
K_- & \geq & R \\
& \geq & \sup_{P \in \mathbb{T}} I(P,Q^*) \\
& \geq & \min_Q \sup_{P \in \mathbb{T}} I(P,Q) \\
& = & K_+ \\
& \geq & K_-,
\end{eqnarray*}
where the first inequality follows from 1), the second from 3), and
the last from Proposition \ref{K-K+relation}. Consequently, all the
inequalities are equalities, $R = K_- = K_+$, and the use of ``min''
in the definition of $K_+$ is justified.

$\Rightarrow$: Since $R = K_- \leq K_+ < \infty$, it follows from
the definition of $K_-$ that there exists a sequence $(\mu_n : n \in
\mathbb{N})$ such that $\lim_n J(\mu_n, \mathbb{T}) = R$.

Now consider the sequence of vectors in $\mathbb{R}^{|\mathbb{X}|
|\mathbb{Y}|}$ given by $F_n = \int_{\mathbb{T}} d\mu_n(P) f(P)$.
This is a sequence of scaled PMFs given by $F_n = d_n \cdot \mu_n
f$, where $d_n$ is given by (\ref{outputPMFNormalizer}). The
sequence resides in a compact space of scaled PMFs and therefore has
a cluster point $F^*$ which can be normalized to get the PMF $Q^*$.
Moreover we can find a subsequence of $(F_n : n \in \mathbb{N})$
such that $\lim_k F_{n_k} = F^*$. We redefine the sequence $\mu_n$
as given by this subsequence, and properties 1) and 2) hold.

Suppose now that there is a $P_0 \in \mathbb{T}$ such that 3) is
violated, {\it i.e.},
\[
I(P_0,Q^*) > K_- .
\]
Consider the convex combinations of measures
\begin{equation}
\label{nuMixtures} \nu_{n, \lambda} = (1 - \lambda) \mu_n +
(\lambda) \delta_{P_0},
\end{equation}
where $\delta_{P_0}$ is the atomic distribution on $P_0$.

From (\ref{nuMixtures}), (\ref{outputPMFUnnormalized}), and
(\ref{J-Fexpansion}), we have
\begin{eqnarray}
s_n(\lambda) & \stackrel{\Delta}{=} & J(\nu_{n, \lambda}, \mathbb{T}) \nonumber \\
& = & \mbox{sign}(\rho) \cdot h \left( (1-\lambda)F_n + \lambda
f(P_0) \right). \nonumber
\end{eqnarray}
Since $\mbox{sign}(\rho)h(\cdot)$ is a concave and therefore
continuous function of its vector-valued argument, $s_n(\lambda)$
converges point-wise to
\[
s(\lambda) = \mbox{sign}(\rho) \cdot h \left( (1-\lambda)F^* +
\lambda f(P_0) \right),
\]
for $\lambda \in [0,1]$. In particular, $s(0) = \lim_n s_n(0) =
K_-$. $s(\lambda)$ is a concave function of $\lambda$ since
$\mbox{sign}(\rho) h(\cdot)$ is concave and the argument is linear
in $\lambda$. Let $\dot{s}(0)$ be the one-sided derivative of
$s(\lambda)$ evaluated at $\lambda=0$ (\emph{i.e.}, limit as
$\lambda \downarrow 0$). We can straightforwardly check that
\[
\dot{s}(0) = I(P_0, Q^*) - K_- > 0,
\]
with the possibility that the value (slope at $\lambda = 0$) may be
$+ \infty$.

We have therefore established that $s(\lambda)$ has $s(0) = K_-$, is
concave and therefore continuous in $[0,1]$, and has strictly
positive slope at $\lambda = 0$. Consequently, $s(\lambda)
> K_-$ for some $0 < \lambda < 1$. Since
\[ J(\nu_{n, \lambda}, \mathbb{T}) = s_n(\lambda) \rightarrow s(\lambda) >
K_-
\]
contradicts the definition of $K_-$, 3) must hold.

To show uniqueness of $Q^*$, suppose there were another $R^*$ and
another sequence of measures $(\pi_n : n \in \mathbb{N})$ satisfying
1), 2) and 3). We can get two cluster points $F^*$ and $G^*$ that
when normalized lead to $Q^*$ and $R^*$, respectively. Then with
$\nu_n = \frac{1}{2} \mu_n + \frac{1}{2} \pi_n$, we have
\begin{eqnarray*}
J \left( \nu_n, \mathbb{T} \right) & \rightarrow & \mbox{sign}(\rho)
\cdot h \left( \frac{1}{2}F^* + \frac{1}{2}G^* \right)
\\
& > & \frac{1}{2} \cdot \mbox{sign}(\rho) \cdot h \left(F^*\right) +
\frac{1}{2} \cdot \mbox{sign}(\rho) \cdot
h \left( G^* \right)\\
& = & \frac{1}{2} K_- + \frac{1}{2} K_- \\
& = & K_-,
\end{eqnarray*}
a contradiction. The strict inequality above is due to strict
concavity of $\mbox{sign}(\rho)h(\cdot)$ when $\rho > -1$ and $\rho
\neq 0$.
\end{proof}
\vspace*{.1in}

\subsection{Proof of Theorem \ref{minSupLalphaTheorem}}
\label{ProofOfExistenceTheorem}
\begin{proof} From (\ref{LIrelationGeneral}), it is clear that
\[
C = \frac{1}{\rho} \log \left( \mbox{sign}(\rho) \cdot K_+ \right).
\]
$Q$ attains the min-sup value $K_+$ in Definition \ref{minSupIf} if
and only if $Q$ attains the min-sup value $C$ in Definition
\ref{minSupLalpha}. Proposition \ref{GallagerExtn} guarantees the
existence and uniqueness of such a $Q$.
\end{proof}

\section{Examples}

\label{examples}

In this section we look at two example families of PMFs, and
identify their $L_{\alpha}$-centers and radii. We focus on guessing
without side information. We also take a closer look at the binary
memoryless channel and obtain tighter upper bounds on redundancy
than those obtained via Theorem \ref{guessingUnderUncertainty}.
Throughout this section, therefore, $0 < \alpha < 1$ and
$|\mathbb{Y}| = 1$. The uncertainty set will thus be PMFs in
$\mathbb{X}$ (with no reference to $|\mathbb{Y}|$).

\subsection{The family of discrete memoryless sources}

\label{DMSExample}

Let $\mathbb{A}$ be a finite alphabet set, $n$ a positive integer,
and $\mathbb{X} = \mathbb{A}^n$. We wish to guess $n$-strings with
letters drawn from $\mathbb{A}$. Let $a^n = (a_1, \cdots, a_n) \in
\mathbb{A}^n$. Let ${\cal{P}}(\mathbb{X})$ denote the set of all
PMFs on $\mathbb{X}$.

Let $\mathbb{T}$ be the set of all discrete memoryless sources (DMS)
on $\mathbb{A}$, {\it i.e.},
\begin{eqnarray*}
\mathbb{T} = \left\{ P_n \in {\cal{P}} \left(\mathbb{A}^n \right)
\mid P_n(a^n) = \prod_{i=1}^n P(a_i), ~ \forall a^n \in
\mathbb{A}^n, \right. \\
\left. \mbox{ and } P \in {\cal{P}}(\mathbb{A}) \right\},
\end{eqnarray*}
The parameters of the source $P$ are unknown to the guesser. Arikan
and Merhav \cite{Arikan-Merhav} provide a guessing scheme for this
uncertainty set. The scheme happens to be independent of $\rho$.
Moreover, their guessing scheme has the same asymptotic performance
as the optimal guessing scheme. Their guessing order proceeds in the
increasing order of empirical entropies; strings with identical
letters are guessed first, then strings with exactly one different
letter, and so on. Within each type of sequence, the order of
guessing is irrelevant. Denote this guessing list by $G_n$. Arikan
and Merhav \cite[Theorem 1]{Arikan-Merhav} showed that for any $P_n
\in \mathbb{T}$,
\[
\lim_{n \rightarrow \infty} \frac{1}{n} R(P_n,G_n) = 0.
\]
The above result is couched in our notation. This indicates that
$\mathbb{T}$, the family of all DMSs on $\mathbb{A}$, is not rich
enough in the sense that there exists a ``universal'' guessing
scheme. The following result makes this notion more precise.

\vspace*{.1in}

\begin{thm} ({\it Family of DMSs on $\mathbb{A}$})
\label{iidSources} Let $m = |\mathbb{A}|$. The $L_{\alpha}$-radius
$C_n$ of the family of discrete memoryless sources on $\mathbb{A}$
satisfies
\[
C_n \leq \frac{m-1}{2} \log \frac{n}{2 \pi} + u_m + \varepsilon_n,
\]
where $u_m = \log \left( \Gamma(1/2)^m / \Gamma(m/2) \right)$, a
constant that depends on the alphabet size, and $\varepsilon_n$ is a
sequence in $n$ that vanishes as $n \rightarrow \infty$. $\hfill
\Box$
\end{thm}

\vspace*{.1in}

\begin{proof}
Recall that $\rho >0$. $P_n$ is the joint PMF of the $n$-string with
individual letter probabilities $P$. Let $P_n \mapsto P'_n$
according to the mapping given in (\ref{tilted}). It is easy to
verify that $P'_n$ is the joint PMF of the $n$-string with
individual letter probabilities $P'$, where $P \mapsto P'$ according
to the mapping (\ref{tilted}), and therefore $P'_n$ also belongs to
$\mathbb{T}$. Furthermore, for a fixed $a^n \in \mathbb{A}^n$, let
$\hat{S}_{a^n}$ be the PMF of letter frequencies in $a^n$, and
define
\[
\hat{S}_{a^n,n}(x^n) \stackrel{\Delta}{=} \prod_{i=1}^n
\hat{S}_{a^n}(x_i),
\]
for every $x^n \in \mathbb{A}^n$. Note that $\hat{S}_{a^n,n}$ is not
necessarily a PMF. Xie and Barron \cite[Theorem 2]{200003TIT_XieBar}
show that there is a PMF on $\mathbb{A}^n$, say $Q'_n$, and a
vanishing sequence $\varepsilon_n$, such that for every discrete
memoryless source $P'_n$, the following holds:
\begin{eqnarray}
\label{XieBarronUpperBoundStep1} \max_{a^n \in \mathbb{A}^n} \log
\frac{P'_n(a^n)}{Q'_n(a^n)} & \leq & \max_{a^n \in
\mathbb{A}^n} \log \frac{\hat{S}_{a^n,n}(a^n)}{Q'_n(a^n)} \\
& \leq & r_n \nonumber \\
\label{XieBarronUpperBound} & \stackrel{\Delta}{=} & \frac{m-1}{2}
\log \frac{n}{2 \pi} + u_m + \varepsilon_n.
\end{eqnarray}
Define the PMF $Q_n$ as follows:
\[
Q_n(\cdot) \propto \left( Q'_n(\cdot) \right)^{1/\alpha},
\]
the inverse of the mapping in (\ref{tilted}). We then have the
following series of inequalities:

\begin{eqnarray}
\lefteqn{ L_{\alpha}(P_n,Q_n) } \nonumber \\
\label{d1} & = & \frac{1}{\rho} \log \left( \sum_{a^n \in
\mathbb{A}^n}
P'_n(a^n) \left( \frac{P'_n(a^n)}{Q'_n(a^n)} \right)^{\rho} \right) \\
\label{d2} & \leq & \frac{1}{\rho} \left( \log \sum_{a^n \in
\mathbb{A}^n} P'_n(a^n) \cdot
\exp \{ \rho r_n \} \right) \\
& = & \frac{1}{\rho} \log \left( \exp \{ \rho r_n \} \right) \nonumber \\
& = & r_n, \nonumber
\end{eqnarray}
where (\ref{d1}) follows from (\ref{LDrelation}) and (\ref{d2}) from
(\ref{XieBarronUpperBound}). Taking the supremum over all $P_n$
yields the theorem.
\end{proof}

\vspace*{.1in}

{\it Remark} : Redundancy in guessing is thus upper bounded by $r_n
+ \log (1 + n \ln |\mathbb{A}|)$. Since the $L_{\alpha}$-radius
grows with $n$ as $O (\log n)$, the normalized redundancy $C_n / n$
vanishes. This implies that we can get a ``universal'' guessing
strategy. Theorem \ref{iidSources} suggests the use of $Q_n$, which
in general may depend on $\rho$. Arikan and Merhav's technique of
guessing in the order of increasing empirical entropy is another
universal guessing technique.

Given any guessing scheme, how do we ``measure'' the set of DMSs
which result in relatively large redundancy? The following theorem
answers this question, and uses a strong version of the redundancy
capacity theorem of universal coding in \cite{199505TIT_MerFed} and
\cite{Feder-Merhav}.

\vspace*{.1in}

\begin{thm}
Let $Q_n$ be any PMF on $\mathbb{A}^n$. Let $\mu$ be a probability
measure on $(\mathbb{T}, {\cal{T}})$ and let $P'_{n,\mu} =
\int_{\mathbb{T}} d \mu(P'_n) ~P'_n$. Then for any DMS $P_n$, we
have
\[
L_{\alpha}(P_n,Q_n) \geq D(P'_n \parallel P'_{n,\mu}) - \lambda_n
\]
except on a set $B$ of $\mu$-probability $\mu \{ B \} \leq 2^{-n
\lambda_n}$.
\end{thm}
\vspace*{.1in}
\begin{proof}
Observe that $\rho > 0$. An application of Jensen's inequality to
the concave function $\log(\cdot)$ yields
\begin{eqnarray*}
L_{\alpha}(P_n,Q_n) & = & \frac{1}{\rho} \log \left( \sum_{a^n \in
\mathbb{A}^n}
P'_n(a^n) \left( \frac{P'_n(a^n)}{Q'_n(a^n)} \right)^{\rho} \right) \\
& \geq & \frac{1}{\rho} \sum_{a^n \in \mathbb{A}^n}
P'_n(a^n) \log \left( \frac{P'_n(a^n)}{Q'_n(a^n)} \right)^{\rho} \\
& = & D(P'_n \parallel Q'_n).
\end{eqnarray*}

The theorem then follows from \cite[Theorem 2]{Feder-Merhav} which
states that the redundancy in source compression $D(P'_n  \parallel
Q'_n)$ is at least as large as $D(P'_n  \parallel  P'_{n,\mu}) -
\lambda_n$ except on a set $B$ of $\mu$-probability upper bounded by
$2^{-n \lambda_n}$.
\end{proof}

\vspace*{.1in}

{\it Remark} : In particular, we may do the following. We choose
$\mu$ such that $D(P'_n  \parallel  P'_{n,\mu}) = r_n$. (This can be
done since the inf-sup value of $\inf_{Q'_n} \sup_{P'_n} D(P'_n
\parallel Q'_n)$ is $r_n$, as remarked in \cite[Remark 5 after
Theorem 2]{200003TIT_XieBar}. We may then choose $\lambda_n$ such
that $n \lambda_n \rightarrow \infty$ so that $2^{-n \lambda_n}$
vanishes with $n$, but $\lambda_n$ is negligibly small compared to
$r_n$. (For example, for the family of DMSs, $r_n = O(\log n)$ and
therefore we may set $\lambda_n = (\log \log n)/(\log n)$). Then,
the set of sources $P$ for which $L_{\alpha}(P_n,Q_n) \leq r_n -
\lambda_n$ has negligible $\mu$-probability for all sufficiently
large $n$. Equivalently, with high $\mu$-probability (at least $1 -
2^{-n \lambda_n}$), $L_{\alpha}(P_n,Q_n) > r_n - \lambda_n$.

Since $L_{\alpha}$ quantifies the redundancy in Campbell's coding
problem to within unity, the above remark leads us to conclude that
the redundancy in that problem is tightly bounded as $\frac{m-1}{2}
(\log n)$ (up to a constant).

In the guessing context, since the nuisance term $\log (1 + n \ln
m)$ grows as $\log n + \log \ln m$ for large $n$, we deduce that
with high $\mu$-probability (at least $1 - 2^{-n \lambda_n} $), the
guessing redundancy of any strategy is at least $r_n - \lambda_n -
\log(1 + n \ln m)$, which for large $n$ is
\begin{equation}
\label{DMSredundancy}
  \frac{m-3}{2} \log n + u_m + \frac{m-1}{2} \log (2 \pi) - \log \ln m
  + \varepsilon_n - \lambda_n.
\end{equation}
This fact and Theorem \ref{iidSources} immediately lead us to
conclude that for $m \geq 4$, the redundancy is between
$\frac{m-3}{2} \log n$ and $\frac{m+1}{2} \log n$ for large $n$
(ignoring constants and smaller order terms). For $m=2$ and $m=3$,
the lower bound in (\ref{DMSredundancy}) is useless, and the upper
bound $\frac{m+1}{2} \log n$ may not be tight. The case of $m=2$ is
addressed in the next subsection. Tighter upper bounds for $m=3$
remain to be found.

\subsection{Guessing an unknown binary memoryless source}

The $L_{\alpha}$-based bounding technique suggested by Theorem
\ref{guessingUnderUncertainty} provides good bounds on guessing
redundancy for large $n$ when the DMS's alphabet size $m \geq 4$. In
this subsection, we identify tighter upper bounds on the guessing
redundancy of a binary memoryless source using a more direct
approach.

Let $\mathbb{A} = \{ 0,1 \}$. There is only one unknown parameter,
\emph{i.e.}, $p = P(1)$. The probability of any $n$-string is given
by
\[
  P_n(x^n) = p^{N(x^n)} (1-p)^{n - N(x^n)} = (1-p)^n \left(
  \frac{p}{1-p} \right)^{N(x^n)},
\]
where $N(x^n)$ is the number of 1s in the string $x^n$. Since
$P_n(x^n)$ is monotonic in $N(x^n)$, it immediately follows that
when $p>1/2$, the optimal guessing order is to guess the string of
all 1s, followed by all strings with exactly one 0, followed by all
strings with exactly two 0s, and so on, \emph{viz.}, in the
decreasing order of number of 1s in the string. Note that the
optimal guessing sequence is the same for all sources whose $p>1/2$.
Exactly the opposite is true when $p<1/2$ - the guessing proceeds in
the increasing order of number of 1s, the first guess being the
all-0 sequence.

Thus there are only two optimal guessing lists for the binary
memoryless source. By guessing one element from each list, skipping
those already guessed, we obtain a guessing list that requires at
most twice the optimal number of guesses, \emph{i.e.}, $G(x^n) \leq
2 G_{P_n}(x^n)$ for every $x^n \in \mathbb{A}^n$. This guessing list
is one of those that proceed in the increasing order of empirical
entropy. Clearly then, the redundancy is upper bounded by the
constant $\log 2$, a bound tighter than Theorem \ref{iidSources}.
$C_n/n$ therefore vanishes as $(\log 2)/n$. It is not known if this
is the tightest upper bound.

\subsection{Arbitrarily varying sources}

\label{AVSExample}

For the family of DMSs, we saw in Section \ref{DMSExample} that the
redundancy is upper bounded by $O(\log n)$. In this section we look
at the example of finite-state arbitrarily varying sources (FS-AVS)
for which the redundancy grows linearly with $n$. Yet again, for
exposition purposes, we assume $|\mathbb{Y}| = 1$.

As before, let $\mathbb{X} = \mathbb{A}^n$. Let $\mathbb{S}$ be a
finite set of {\it states}, and for each $s \in \mathbb{S}$, let $P
( \cdot \mid s)$ be a PMF on the finite set $\mathbb{A}$. An {\it
arbitrarily varying source} (AVS) is a sequence of
$\mathbb{A}$-valued random variables $X_1, X_2, \cdots$, such that
$X_i$'s are independent and the probability of an $n$-string $x^n$
is governed by an arbitrary state sequence $s^n \in \mathbb{S}^n$ as
follows:
\[
P_n(x^n \mid s^n) = \prod_{i=1}^n P(x_i \mid s_i).
\]

Observe that for a fixed $n$, there are only $|\mathbb{S}|^n$
sources in the uncertainty set. Let $T_{s^n}$ be the subset of all
sequences in $\mathbb{S}^n$ with the same letter-frequencies as
$s^n$. $T_{s^n}$ is also referred to as the type of the sequence
$s^n$ \cite{CK-IT81}. If the letter frequencies are given by a PMF
$U$ on $\mathbb{S}$, we refer to $T_U$ as the type of sequences. Let
$V$ be a stochastic matrix given by $V(x \mid s)$ for $x \in
\mathbb{A}$ and $s \in \mathbb{S}$. Then for a particular sequence
$s^n$, we refer to $T_V(s^n)$, the set of sequences that are of
conditional type $V$ given $s^n$, as the $V$-{\it shell} of $s^n$.

\vspace*{.1in}

\begin{proposition}
\label{GuessingAVSTypeProposition} Let $0 < \alpha < 1$. Let $T_U$
be a type of sequences on $\mathbb{S}^n$. Let the uncertainty set
$\mathbb{T}$ be given by $\mathbb{T} = \{ P_n(\cdot \mid s^n) \mid
s^n \in T_U \}$. The $L_{\alpha}$-radius of this family is given by
\begin{equation}
\label{redundancyEquation} R_n(T_U) \stackrel{\Delta}{=}
H_{\alpha}(Q^*_n) - \frac{1}{|T_U|} \sum_{s^n \in T_U}
H_{\alpha}(P_n(\cdot \mid s^n)),
\end{equation}
where the $L_{\alpha}$-center $Q^*_n$ is given by
\begin{equation}
\label{minimizerAVS} Q^{*}_n(\cdot) = \frac{1}{|T_U|} \sum_{s^n \in
T_U} P_n(\cdot \mid s^n).
\end{equation}
$ \hfill \Box$
\end{proposition}
\vspace*{.1in}

{\it Remarks} : 1) It will be apparent from the proof that the
quantity $H_{\alpha}(P_n(\cdot \mid s^n))$ in
(\ref{redundancyEquation}) depends on $s^n$ only through its type,
and hence the average over all sequences in the type may be replaced
by the value for any specific $s^n \in T_U$.

2) All PMFs in the uncertainty set are spaced equally apart (in the
sense of $L_{\alpha}$-divergence) from the $L_{\alpha}$-center
$Q^*_n$.

3) Guessing in the decreasing order of $Q^*_n$-probabilities results
in a redundancy in guessing that is upper bounded by $R_n(T_U) +
\log(1 + n \ln |\mathbb{A}|)$.

4) $\mbox{sign}(\rho) \cdot h(P)$ is a concave function of $P$. It
follows from (\ref{conditionalRenyiEntropy}) that $H_{\alpha}(P)$ is
also a concave function of $P$ for $0 < \alpha < 1$. By Jensen's
inequality, $R_n(T_U) \geq 0$. (For $\alpha > 1$, $H_{\alpha}(P)$ is
neither concave nor convex in $P$).

5) For any guessing strategy, there exists at least one sequence
$s^n \in T_U$ for which the redundancy is lower bounded by $R_n(T_U)
- \log(1 + n \ln |\mathbb{A}|)$. We will see later in Proposition
\ref{AVSConvergence} that if the $U$ sequence (parameterized by $n$)
converges as $n \rightarrow \infty$ to a PMF $U^* \in
{\cal{P}}(\mathbb{S})$, then $\frac{1}{n}R_n(T_U)$ converges to a
strictly positive constant. Thus $R_n(T_U)$ grows linearly with $n$,
thereby making the converse meaningful; the nuisance term
$\log(1+n\ln |\mathbb{A}|)$ grows only logarithmically in $n$.

\vspace*{.1in}
\begin{proof}
Note that given an $n$, the uncertainty set is finite. We will
simply show that the candidate $L_{\alpha}$-center satisfies the
necessary and sufficient condition (\ref{NASC}) given in Section
\ref{FiniteCaseNASC}. From (\ref{IDefinitionRho}), it is sufficient
to show that
\begin{eqnarray}
\lefteqn{ I_f(P'_n(\cdot \mid s^n \parallel Q^{*'}_n) } \nonumber \\
\label{e1} & = & \frac{\sum_{x^n \in \mathbb{A}^n}
P_n(x^n \mid s^n) \left( Q^{*'}_n(x^n) \right)^{-\rho}}{h(P_n(\cdot \mid s^n))} \\
& = & K, \nonumber
\end{eqnarray}
where $K$ is some constant that depends only on $n$ and $T_U$. We
will show that the numerator and denominator in (\ref{e1}) do not
depend on the actual $s^n$, so long as $s^n \in T_U$.

Observe that the stochastic matrix that defines the conditional PMF
is given by $P(x \mid s)$ for $x \in \mathbb{A}$ and $s \in
\mathbb{S}$. Consider $h(P_n(\cdot \mid s^n))$. First
\begin{eqnarray*}
\lefteqn{ \sum_{x^n \in \mathbb{A}^n} \left( P_n(x^n \mid s^n) \right)^{\alpha} } \\
& = & \sum_{V} |T_V(s^n)| \exp \left\{-n \alpha \left[ D(V \parallel
P \mid U) + H(V \mid U) \right] \right\}
\end{eqnarray*}
where the sum is over all conditional types $V$. All the quantities
inside the summation, including $|T_V(s^n)|$, depend on $s^n$ only
through $T_U$, and therefore $h(P_n (\cdot \mid s^n))$ depends on
$s^n$ only through $T_U$.

Next, $Q^{*}_n(x^n)$ depends on $x^n$ only through $T_{x^n}$. This
is easily seen via a permutation argument. Given two
$\mathbb{A}$-sequences of the same type, let $\pi$ be a permutation
that takes $(x^n, s^n)$ to $((x_{\pi(1)}, \cdots, x_{\pi(n)}),
(s_{\pi(1)}, \cdots, s_{\pi(n)}))$, where $s^n$ and $(s_{\pi(1)},
\cdots, s_{\pi(n)})$ are the two given $\mathbb{A}$-sequences. This
permutation $\pi$ leaves $P_n(x^n \mid s^n)$ unchanged. Moreover,
the sum continues to be over
\begin{eqnarray*}
T_U =  \left\{ (s_{\pi(1)}, s_{\pi(2)}, \cdots , s_{\pi(n)}) \in
\mathbb{S}^n \mid \right. \\
\left. s^n = (s_1, \cdots, s_n) \in T_U \right\}.
\end{eqnarray*}
Thus $Q^{*}_n(x^n)$ and therefore $Q^{*'}_n(x^n)$ depend on $x^n$
only through $T_{x^n}$.

Finally, given two $\mathbb{A}$-sequences of the same type $T_U$,
the above permutation argument indicates that
\begin{eqnarray*}
\sum_{x^n \in \mathbb{A}^n} P_n(x^n \mid s^n) \left( Q^{*'}_n(x^n)
\right)^{-\rho},
\end{eqnarray*}
the numerator of (\ref{e1}), depends on $s^n$ only through $T_U$.

That $R_n(T_U)$ is given by (\ref{redundancyEquation}) follows from
(\ref{modify-w}), (\ref{normalizer}),
(\ref{minMaxLalphaExpression}), the fact that $h(P_n(\cdot \mid
s^n))$ is a constant over all $s^n \in T_U$, and
(\ref{conditionalRenyiEntropy}). This concludes the proof.
\end{proof}
\vspace*{.1in}

The number of different types of sequences grows polynomially in
$n$, in particular, this number is upper bounded by
$(n+1)^{|\mathbb{S}|}$. We can use this fact to stitch together the
guessing lists for the different types of sequences on
$\mathbb{S}^n$ and get one list that does only marginally worse than
the list obtained by knowing the type of the state sequence.

\vspace*{.1in}
\begin{proposition}
Let $0 < \alpha < 1$. Let the uncertainty set $\mathbb{T}$ be given
by $\mathbb{T} = \{ P_n(\cdot \mid s^n) \mid s^n \in \mathbb{S}^n
\}$. There is a guessing strategy such that for every $T_U$, the
redundancy is upper bounded by
\[
R_n(T_U) + \log (1 + n \ln |\mathbb{A}|) + |\mathbb{S}| \log(n + 1).
\]
whenever $s^n \in T_U$. $ \hfill \Box$
\end{proposition}
\vspace*{.1in}

\begin{proof}
Let $N$ be the number of types. $N$ is upper bounded by
$(n+1)^{|\mathbb{S}|}$. Fix an arbitrary order on these types. Let
the $k$th type be $T_U$. Set $G_k = G_{T_U}$, where $G_{T_U}$ is the
guessing strategy that is obtained knowing that $s^n \in T_U$, via
Proposition \ref{GuessingAVSTypeProposition}. It proceeds in the
decreasing order of probabilities of the $L_{\alpha}$-center of the
uncertainty set indexed by $T_U$.

We now stitch together the guessing lists $G_1, G_2, \cdots, G_N$ to
get a new guessing list $G$, as follows. Think of $G_k$ as a column
vector of size $|\mathbb{A}^n| \times 1$ and let $H$ be the column
vector of size $N \cdot |\mathbb{A}^n| \times 1$ obtained by reading
the entries of the matrix $\left[G_1 ~ G_2 ~ \cdots ~ G_N \right]$
in raster order (one row after another). Every $\mathbb{A}$ would
have figured exactly once in the $G_k$ list, and therefore occurs
exactly $N$ times in the $H$ list. Next, prune the $H$ list. For
each $i$, if there exists an index $j$ with $j < i$ and $H_i = H_j$,
set $H_i = \delta$. This indicates that the $i$th string already
figures in the final guessing list. Finally remove all $\delta$'s to
obtain the desired guessing list $G : \mathbb{A}^n \rightarrow \{1,
2, \cdots, |\mathbb{A}|^n \}$, where $G(x^n)$ is the unique position
at which $x^n$ occurs in the pruned $H$ list.

Clearly, for every $x^n$ and for every $k$ such that $1 \leq k \leq
N$, we have $G(x^n) \leq N G_k(x^n)$. Indeed, $x^n$ occurs in the
position $(G_k(x^n), k)$ in the matrix constructed above. It
therefore occurs in position $(G_k(x^n)-1)N + k$ and therefore
before the position $N G_k(x^n)$ in the unpruned $H$ list. It cannot
be placed any later in the pruned $H$ list, and thus $G(x^n) \leq N
G_k(x^n)$.

The above observation leads to
\[
\frac{1}{\rho} \log \mathbb{E} \left[ G(X^n)^{\rho} \right] \leq
\frac{1}{\rho} \log \mathbb{E} \left[ G(X^n)^{\rho} \right] + \log
N.
\]
The proposition follows from Theorem
\ref{guessingRedundancyTheorem}, Proposition
\ref{GuessingAVSTypeProposition}, and the bounding $N \leq
(n+1)^{|\mathbb{S}|}$.
\end{proof}
\vspace*{.1in}

We finally remark that the min-sup redundancy for the finite-state
arbitrarily varying source grows linearly with $n$ under some
circumstances.

\vspace*{.1in}
\begin{proposition}
\label{AVSConvergence} For a fixed $n$, let $U$ be a PMF on
$\mathbb{S}$ and $T_U$ the corresponding type. Let the sequence $U$
(as a function of $n$) converge to a PMF $U^* \in
{\cal{P}}(\mathbb{S})$ as $n \rightarrow \infty$. Then
\[
\lim_n \frac{1}{n} R_n(T_U) = R,
\]
where $R \geq 0$. $\hfill \Box$
\end{proposition}

\vspace*{.1in}

\begin{proof}
The second term in the right-hand side of
(\ref{redundancyEquation}), after normalization by $n$, converges to
a nonnegative real number as seen below:
\begin{eqnarray}
\lefteqn{ \frac{1}{n} H_{\alpha}(P_n(\cdot \mid s^n)) } \nonumber \\
& = & \frac{1}{n(1-\alpha)} \log \sum_{x^n \in \mathbb{A}^n}
\prod_{i=1}^n
P(x_i \mid s_i)^{\alpha} \nonumber \\
& = & \frac{1}{n(1-\alpha)} \log \prod_{s \in \mathbb{S}} \left(
\sum_{x \in \mathbb{A}} P(x \mid s)^{\alpha} \right)^{n U(s)} \nonumber \\
& = & \sum_{s \in \mathbb{S}} U(s) H_{\alpha}(P(\cdot \mid s)) \nonumber \\
\label{secondTermOfAVS} & \rightarrow & \sum_{s \in \mathbb{S}}
U^*(s) H_{\alpha}(P(\cdot \mid s)).
\end{eqnarray}

We next consider the first term on the right-hand side of
(\ref{redundancyEquation}) after normalization, {\em i.e.},
$H_{\alpha}(Q_n^*) / n$, where $Q_n^*$ is given by
(\ref{minimizerAVS}).

\vspace*{.1in}

\begin{lemma}
\label{AVSConvergenceLemma} For a fixed $n$, let $U$ be a PMF on
$\mathbb{S}$ and $T_U$ the corresponding type. Let the sequence $U$
(as a function of $n$) converge to a PMF $U^* \in
{\cal{P}}(\mathbb{S})$ as $n \rightarrow \infty$. Let $V$ be the
output PMF when the input PMF on $\mathbb{S}$ is $U$ and the channel
is $P$. Furthermore, let $V^*$ be the limiting output PMF as $n
\rightarrow \infty$. Then $\lim_n \frac{1}{n} H_{\alpha}(Q_n^*) =
H_{\alpha}(V^*)$. $\hfill \Box$
\end{lemma}

\vspace*{.1in}

As a consequence of this lemma and (\ref{secondTermOfAVS}), we have
\[
\frac{1}{n} R_n(T_U) \rightarrow H_{\alpha}(V^*) - \sum_{s \in
\mathbb{S}} U^*(s) H_{\alpha}(P(\cdot \mid s)) \stackrel{\Delta}{=}
R.
\]
By the strict concavity of $H_{\alpha}(\cdot)$ for $0 < \alpha < 1$,
and Jensen's inequality, we have $R \geq 0$. This concludes the
proof of the theorem.
\end{proof}
\vspace*{.1in}

{\it Remarks} : $R=0$ if and only if either $(i)$ $U(s) = 1$ for
some $s \in \mathbb{S}$, or $(ii)$ $P(\cdot \mid s)$ does not depend
on $s$, {\em i.e.}, the state does not affect the source. Thus, for
all but the trivial finite-state arbitrarily varying sources, the
min-sup redundancy grows exponentially with $n$ at a rate $R$. This
means that the guessing strategy that achieves the min-sup
redundancy has an exponential growth rate strictly bigger than that
of the best strategy obtained with knowledge of the state sequence.

\vspace*{.1in}

We now prove the rather technical Lemma \ref{AVSConvergenceLemma}.

\vspace*{.1in}

\begin{proof}

(a) We first show that $\lim_n \frac{1}{n} H_{\alpha}(Q_n^*) \leq
H_{\alpha}(V^*)$. Let $U_n$ be the PMF on $\mathbb{S}^n$ given by
$U_n(s^n) = \prod_{i=1}^n U(s_i)$. Let $U_n \{ T \}$ denote the
$U_n$-probability of the set $T$. From (\ref{minimizerAVS}), we may
write
\begin{eqnarray}
  \lefteqn{ \sum_{x^n \in \mathbb{A}^n} Q_n^*(x^n)^{\alpha} }
  \nonumber \\
  &=& \sum_{x^n \in \mathbb{A}^n} \left( \frac{1}{|T_U|} \sum_{s^n \in T_U} P_n(x^n \mid s^n)\right)^{\alpha} \nonumber \\
  &=& \sum_{x^n \in \mathbb{A}^n} \left( \frac{1}{U_n \{ T_U \}} \frac{U_n \{ T_U \} }{|T_U|} \sum_{s^n \in T_U} P_n(x^n \mid s^n)\right)^{\alpha} \nonumber \\
  \label{f1}
  &=& \frac{1}{U_n \{ T_U \}^{\alpha}} \sum_{x^n \in \mathbb{A}^n} \left( \sum_{s^n \in T_U} U_n(s^n) P_n(x^n \mid
  s^n)\right)^{\alpha} \\
  \label{f2}
  &\leq& (n + 1)^{|\mathbb{S}| \alpha} \sum_{x^n \in \mathbb{A}^n} \left( \sum_{s^n \in \mathbb{S}^n} U_n(s^n) P_n(x^n \mid
  s^n)\right)^{\alpha} \\
  &=& (n + 1)^{|\mathbb{S}| \alpha} \sum_{x^n \in \mathbb{A}^n} V_n(x^n)^{\alpha} \nonumber \\
  \label{f3}
  &=& (n + 1)^{|\mathbb{S}| \alpha} \left( \sum_{x \in \mathbb{A}}
  V(x)^{\alpha} \right)^n,
\end{eqnarray}
where (\ref{f1}) follows from the observation that $U_n(s^n) = U_n
\{ T_U \} / |T_U|$ for all $s^n \in T_U$, (\ref{f2}) from  $U_n \{
T_U \} \geq (n + 1)^{-|\mathbb{S}|}$ (see proof of \cite[Lemma
2.3]{CK-IT81}) and by enlarging the sum over $T_U$ to all of
$\mathbb{S}^n$.

From (\ref{f3}) and (\ref{conditionalRenyiEntropy}), we have
\begin{eqnarray*}
  \frac{1}{n} H_{\alpha}(Q_n^*) &\leq& \frac{\alpha |\mathbb{S}|}{1 - \alpha} \frac{\log (n+1)}{n} + H_{\alpha}(V) \\
  &\rightarrow& H_{\alpha}(V^*).
\end{eqnarray*}

\vspace*{.1in}

(b) We now show that $\lim_n \frac{1}{n} H_{\alpha}(Q_n^*) \geq
H_{\alpha}(V^*)$. For a given PMF $U$ on $\mathbb{S}$ and
conditional PMF $P$, let $V$ be the induced PMF on $\mathbb{X}$ and
$W$ the reverse conditional PMF, {\em i.e.}, $W(s \mid x)$ is the
probability of a state $s$ given $x$.

Continuing from (\ref{f1}), we may write
\begin{eqnarray}
  \lefteqn{ \sum_{x^n \in \mathbb{A}^n} Q_n^*(x^n)^{\alpha} } \nonumber \\
  &=& \frac{1}{U_n \{ T_U \}^{\alpha}} \sum_{x^n \in \mathbb{A}^n} \left( \sum_{s^n \in T_U} U_n(s^n) P_n(x^n \mid
  s^n) \right)^{\alpha} \nonumber \\
  \label{g1}
  &\geq& \sum_{x^n \in T_{\overline{Q}}} \left( \sum_{s^n \in T_U} U_n(s^n) P_n(x^n \mid
  s^n) \right)^{\alpha} \\
  \label{g2}
  &\geq& \sum_{x^n \in T_{\overline{Q}}} \left( \sum_{s^n \in T_{\overline{W}}(x^n) \subset T_U} V_n(x^n) W_n(s^n \mid
  x^n) \right)^{\alpha} \\
  \label{g3}
  &=& \sum_{x^n \in T_{\overline{Q}}} \left( V_n(x^n) W_n \left\{ T_{\overline{W}}(x^n) \mid x^n
  \right\} \right)^{\alpha},
\end{eqnarray}
where (\ref{g1}) follows because $U_n\{ T_U \}^{\alpha} \leq 1$ and
the sum over $\mathbb{A}^n$ is restricted to a sum over a type
$T_{\overline{Q}}$ to be chosen later; (\ref{g2}) follows because
$U_n(s^n) P_n( x^n \mid s^n) = V_n(x^n) W_n(s^n \mid x^n)$ and the
sum over $s^n$ is now restricted over a non-void
$\overline{W}$-shell of $x^n$, where $\overline{W}$ will be
appropriately chosen later.

We next observe that for $x^n \in T_{\overline{Q}}$, the following
hold:
\begin{eqnarray*}
  V_n(x^n) & = & 2^{-n \left( H \left( \overline{Q} \right) + D \left( \overline{Q} \parallel V \right) \right)}, \\
  W_n \left\{ T_{\overline{W}}(x^n) \mid x^n \right\} & \geq & (n + 1)^{-|\mathbb{S}| |\mathbb{X}|} \cdot 2^{-n D \left( \overline{W} \parallel W
\mid \overline{Q} \right) },\\
  |T_{\overline{Q}}| & \geq & (n + 1)^{-|\mathbb{X}|} \cdot 2^{n H(\overline{Q})}.
\end{eqnarray*}
Substitution of these inequalities into (\ref{g3}) yields
\begin{eqnarray*}
\lefteqn{ \sum_{x^n \in \mathbb{A}^n} Q_n^*(x^n)^{\alpha} }
\nonumber \\
&& \geq (n+1)^{-|\mathbb{X}| (1 + \alpha |\mathbb{S}|)} \nonumber \\
&&  ~~ \cdot ~ 2^{n \left[ (1 - \alpha) H(\overline{Q}) - \alpha
\left( D \left( \overline{Q}
\parallel V \right)  +  D \left( \overline{W} \parallel W
\mid \overline{Q} \right) \right) \right] }
\end{eqnarray*}
and therefore
\begin{eqnarray}
\lefteqn{ \frac{1}{n} H_{\alpha}(Q_n^*) } \nonumber \\
& \geq & H(\overline{Q}) - \frac{1}{\rho} \left[ D \left(
\overline{Q}
\parallel V \right)  +  D \left( \overline{W} \parallel W
\mid \overline{Q} \right) \right] \nonumber \\
\label{lowerBoundOnNormalizedRenyiEntropyAVS} && - ~
\frac{|\mathbb{X}|(1+\alpha |\mathbb{S}|)}{1-\alpha} \frac{\log
(n+1)}{n}
\end{eqnarray}
for any type $\overline{Q}$ of sequences and for any $\overline{W}$
such that $T_{\overline{W}}(x^n) \subset T_U$ is a non-void shell
for an $x^n \in T_{\overline{Q}}$.

Clearly, the last term in
(\ref{lowerBoundOnNormalizedRenyiEntropyAVS}) vanishes as $n
\rightarrow \infty$.

If we can choose $\overline{Q} = V'$ and $\overline{W} = W$, we will
be done since $H_{\alpha}(V) = H(V') - \frac{1}{\rho} D(V'
\parallel V)$. We cannot do this if $V'$ is not a type of sequences,
or if $W$ is not a conditional type given an $x^n$. But we will show
that as $n \rightarrow \infty$, we can get close enough. The
following arguments make this idea precise.

Define
\[
\delta \stackrel{\Delta}{=} \min \{ W(s \mid x) \mid W(s \mid x)
> 0, ~ s \in \mathbb{S}, x \in \mathbb{X} \}
\]
and consider $D( \overline{W}( \cdot \mid x ) \parallel W(\cdot \mid
x))$. We may restrict our choice of $\overline{W}$ to those that are
absolutely continuous with respect to $W$, {\it i.e.}, $W( \cdot
\mid x) \ll W(\cdot \mid x)$ for every $x \in \mathbb{X}$. For
sufficiently large $n$, we can choose such a $\overline{W}$ that in
addition satisfies
\[
\sum_{s \in \mathbb{S}} \left| W(s \mid x) - \overline{W}(s \mid x)
\right| \leq \varepsilon_n \leq \frac{1}{2}, ~ \forall x \in
\mathbb{X},
\]
and $\varepsilon_n \rightarrow 0$.

We then have
\begin{eqnarray}
\lefteqn { D( \overline{W}( \cdot \mid x ) \parallel W(\cdot \mid x)) } \nonumber \\
& = & H(W(\cdot \mid x)) - H(\overline{W}(\cdot \mid x)) \nonumber \\
&& + ~ \sum_{s \in \mathbb{S}} \left( W(s \mid x) - \overline{W}(s
\mid x) \right) \log
W(s \mid x) \nonumber \\
& \leq & \left| H(W(\cdot \mid x)) - H(\overline{W}(\cdot \mid x))
\right| \nonumber \\
&& - ~ (\log \delta) \sum_{s \in \mathbb{S}} \left| W(s \mid x) -
\overline{W}(s \mid x) \right| \nonumber \\
\label{h1} & \leq & - \varepsilon_n \log
\frac{\varepsilon_n}{|\mathbb{S}|} - \varepsilon_n \log \delta,
\end{eqnarray}
where (\ref{h1}) follows from \cite[Lemma 2.7]{CK-IT81}. After
averaging, we get
\[
D(\overline{W} \parallel W \mid \overline{Q}) \leq - \varepsilon_n
\log \frac{\varepsilon_n}{|\mathbb{S}|} - \varepsilon_n \log \delta
\rightarrow 0.
\]

A similar argument shows that
\begin{eqnarray*}
\lefteqn{ H(\overline{Q}) - \frac{1}{\rho} D \left( \overline{Q}
\parallel V \right) } \\
& = & H_{\alpha}(V) + \left[ H(\overline{Q}) - H(V') \right] \\
~~~ & & - ~ \frac{1}{\rho} \left[ D \left( \overline{Q}
\parallel V \right) - D \left( V' \parallel V \right) \right] \\
& \rightarrow & H_{\alpha}(V^*),
\end{eqnarray*}
where we have made use of the fact that $H_{\alpha}(V) = H(V') -
(1/\rho) D(V' \parallel V)$. This concludes the proof of Lemma
\ref{AVSConvergenceLemma}
\end{proof}

\section{$L_{\alpha}$-projection}

\label{LalphaProjection}

In this section we look at an interesting geometric property of
$L_{\alpha}$ divergence that makes it behave like squared Euclidean
distance, analogous to the Kullback-Leibler divergence. Throughout
this section, we assume $\alpha > -1$ and $\alpha \neq 0$.

We proceed along the lines of \cite{Csiszar-I-proj}. Let
$\mathbb{X}$ and $\mathbb{Y}$ be finite alphabet sets. Let
${\cal{P}}(\mathbb{X} \times \mathbb{Y})$ denote the set of PMFs on
$\mathbb{X} \times \mathbb{Y}$. Given a PMF $R$ on $\mathbb{X}
\times \mathbb{Y}$, the set
\[
B(R, r) \stackrel{\Delta}{=} \left\{ P \in {\cal{P}}(\mathbb{X}
\times \mathbb{Y}) \mid L_{\alpha}(P,R) < r \right\}, ~~~ 0 < r \leq
\infty,
\]
is called an $L_{\alpha}$-sphere (or ball) with center $R$ and
radius $r$. The term ``sphere'' conjures the image of a convex set.
That the set is indeed convex needs a proof since $L_{\alpha}(P,R)$
is not convex in its arguments.

\vspace*{.1in}

\begin{proposition} \label{ballIsConvex}
$B(R,r)$ is a convex set. $\hfill \Box$
\end{proposition}

\vspace*{.1in}

\begin{proof}
Let $P_i \in B(R,r)$ for $i=0,1$. For any $\lambda \in [0,1]$, we
need to show that $P_{\lambda} = (1-\lambda) P_0 + \lambda P_1 \in
B(R,r)$. With $\alpha = 1/(1+\rho)$, and $t = \mbox{sign}(\rho)
\cdot 2^{\rho r}$, we get from (\ref{LIrelationGeneral}) that
\begin{equation}
\label{perPMFub} I(P_i , R) < t, ~~~ i=0,1.
\end{equation}
The proof will be complete if we can show that $I(P_{\lambda} , R) <
t$. To this end,
\begin{eqnarray}
\lefteqn { I(P_{\lambda} , R) } \nonumber \\
& = & \frac{\mbox{sign}(1-\alpha)}{h(P_{\lambda})} \cdot \sum_{y \in
\mathbb{Y}} \sum_{x \in {\mathbb{X}}}
P_{\lambda}(x,y) \left(R'(x \mid y)\right)^{\frac{\alpha-1}{\alpha}}  \nonumber \\
& = & \frac{\mbox{sign}(1-\alpha) }{h(P_{\lambda})} \cdot (1 -
\lambda) \sum_{y \in \mathbb{Y}} \sum_{x \in {\mathbb{X}}} P_0(x,y)
\left(R'(x \mid y)\right)^{\frac{\alpha-1}{\alpha}} \nonumber \\
& & + ~~ \frac{\mbox{sign}(1-\alpha)  }{h(P_{\lambda})}  \cdot
(\lambda) \sum_{y \in \mathbb{Y}} \sum_{x \in {\mathbb{X}}} P_1(x,y)
\left(R'(x \mid y)\right)^{\frac{\alpha-1}{\alpha}} \nonumber \\
\label{a0} & = & \frac{(1 - \lambda) h(P_0) I(P_0, R) + \lambda
h(P_1) I(P_1, R)} {h(P_{\lambda})} \\
\label{a1} & < & t \frac{(1 - \lambda) h(P_0) + \lambda
h(P_1) } {h(P_{\lambda})} \\
& = & |t| \frac{(1 - \lambda) \cdot \mbox{sign}(1 - \alpha) h(P_0) +
\lambda \cdot \mbox{sign}(1 - \alpha)
h(P_1)} {h(P_{\lambda})} \nonumber \\
\label{a2} & \leq & |t| \frac{ \mbox{sign} (1-\alpha)
h(P_{\lambda})}{h(P_{\lambda})} \\
& = & t; \nonumber
\end{eqnarray}
where (\ref{a0}) follows from (\ref{IDefinitionAlpha}), (\ref{a1})
from (\ref{perPMFub}), and (\ref{a2}) from the concavity of
$\mbox{sign}(1 - \alpha) h$.
\end{proof}

\vspace*{.1in}

Proposition \ref{ballIsConvex} shows that $L_{\alpha}(P,R)$ is a
\emph{quasiconvex} function of $P$, its first argument.

When we talk of closed sets, we refer to the usual Euclidean metric
on $\mathbb{R}^{|\mathbb{X}| |\mathbb{Y}|}$. The set of PMFs on
$\mathbb{X} \times \mathbb{Y}$ is closed and bounded (and therefore
compact).

If ${\cal{E}}$ is a closed and convex set of PMFs on $\mathbb{X}
\times \mathbb{Y}$ intersecting $B(R, \infty)$, {\it i.e.} there
exists a PMF $P$ such that $L_{\alpha}(P,R) < \infty$, then a PMF $Q
\in {\cal{E}}$ satisfying
\[
L_{\alpha}(Q,R) = \min_{P \in {\cal{E}}} L_{\alpha}(P,R),
\]
is called the $L_{\alpha}$-{\it projection} of $R$ on ${\cal{E}}$.

\vspace*{.1in}
\begin{proposition}
\label{existenceOfProjection} {\em (Existence of
$L_{\alpha}$-projection)} Let ${\cal{E}}$ be a closed and convex set
of PMFs on $\mathbb{X} \times \mathbb{Y}$. If $B(R,\infty) \cap
{\cal{E}}$ is nonempty, then $R$ has an $L_{\alpha}$-projection on
${\cal{E}}$.
\end{proposition}
\vspace*{.1in}

\begin{proof}
Pick a sequence $P_n \in {\cal{E}}$ with $L_{\alpha}(P_n, R) <
\infty$ such that $L_{\alpha}(P_n, R) \rightarrow \inf_{P \in
{\cal{E}}} L_{\alpha}(P, R)$. This sequence being in the compact
space ${\cal{E}}$ has a cluster point $Q$ and a subsequence
converging to $Q$. We can simply focus on this subsequence and
therefore assume that $P_n \rightarrow Q$ and $L_{\alpha}(P_n, R)
\rightarrow \inf_{P \in {\cal{E}}} L_{\alpha}(P, R)$. $\cal{E}$ is
closed and hence $Q \in {\cal{E}}$. The continuity of the logarithm
function, wherever it is finite, and the condition $L_{\alpha}(P_n,
R) < \infty$ imply that
\begin{eqnarray}
\lim_n L_{\alpha}(P_n, R) & = & \frac{1}{\rho} \log \left(
\mbox{sign}(\rho) \cdot \lim_n I(P_n, R) \right) \nonumber \\
\label{b1} & = & \frac{1}{\rho} \log \left( \mbox{sign}(\rho) \cdot
I(Q, R)
\right) \\
& = & L_{\alpha}(Q, R), \nonumber
\end{eqnarray}
where (\ref{b1}) follows from the observation that
(\ref{IDefinitionAlpha}) is the ratio of a continuous linear
function of $P$ and the continuous concave function
$\mbox{sign}(1-\alpha)h$ that is bounded, and moreover bounded away
from 0.

From the uniqueness of limits we have that $L_{\alpha}(Q, R) =
\inf_{P \in {\cal{E}}} L_{\alpha}(P, R)$. $Q$ is then an
$L_{\alpha}$-projection of $R$ on ${\cal{E}}$.
\end{proof}
\vspace*{.1in}

We next state generalizations of \cite[Lemma 2.1, Theorem
2.2]{Csiszar-I-proj} which show that $L_{\alpha}(P,Q)$ plays the
role of squared Euclidean distance (analogous to the
Kullback-Leibler divergence).

\vspace*{.1in}
\begin{proposition}
\label{pythagoreanProposition} Let $0 < \alpha < \infty, \alpha \neq
1$.
\begin{enumerate}
\item Let $L_{\alpha}(Q,R)$ and $L_{\alpha}(P,R)$ be finite. The
segment joining $P$ and $Q$ does not intersect the
$L_{\alpha}$-sphere $B(R,r)$ with radius $r=L_{\alpha}(Q,R)$, {\em
i.e.},
\[
L_{\alpha}(P_{\lambda},R) \geq L_{\alpha}(Q,R)
\]
for each
\[
P_{\lambda} = \lambda P + (1 - \lambda) Q, ~~~~ 0 \leq \lambda \leq
1,
\]
if and only if
\begin{equation}
\label{pythagoreanPropositionInequality} L_{\alpha}(P,R) \geq
L_{\alpha}(P,Q) + L_{\alpha}(Q,R).
\end{equation}

\item {\em (Tangent hyperplane)} Let
\begin{equation}
\label{innerPMF} Q = \lambda P + (1 - \lambda) S, ~~~~ 0 < \lambda <
1.
\end{equation}
Let $L_{\alpha}(Q,R)$, $L_{\alpha}(P,R)$, and $L_{\alpha}(S,R)$  be
finite. The segment joining $P$ and $S$ does not intersect $B(R,r)$
(with $r = L_{\alpha}(Q, R)$) if and only if
\begin{equation}
\label{pythagoreanPropositionEquality} L_{\alpha}(P,R) =
L_{\alpha}(P,Q) + L_{\alpha}(Q,R).
\end{equation}
$\hfill \Box$
\end{enumerate}
\end{proposition}
\vspace*{.1in}

{\em Remarks}: 1) Under the hypotheses in Proposition
\ref{pythagoreanProposition}.1, we deduce that $L_{\alpha}(P,Q) <
\infty$ as a consequence.

2) The condition (\ref{innerPMF}) implies that $P \leq
\lambda^{-1}Q$ (\emph{i.e.}, every component satisfies the
inequality), and therefore $\mbox{supp}(P) \subset \mbox{supp}(Q)$.
If $0 < \alpha <1$, and $L_{\alpha}(Q,R) < \infty$, then we have
$\mbox{supp}(P) \subset \mbox{supp}(Q) \subset \mbox{supp}(R)$. Thus
both $L_{\alpha}(P,R)$ and $L_{\alpha}(P,Q)$ are necessarily finite.
For $\alpha \in (0,1)$, the requirement that $L_{\alpha}(P,R)$ be
finite can therefore be removed. The requirement is however needed
for $1 < \alpha < \infty$ because even though $\mbox{supp}(P)
\subset \mbox{supp}(Q)$ and $\mbox{supp}(Q) \cap \mbox{supp}(R) \neq
\emptyset$, we may have $\mbox{supp}(P) \cap \mbox{supp}(R) =
\emptyset$ leading to $L_{\alpha}(P,R) = \infty$.

3) Proposition \ref{pythagoreanProposition}.2 extends the analog of
Pythagoras theorem, known to hold for the Kullback-Leibler
divergence, to the family $L_{\alpha}$ parameterized by $\alpha >
0$.

4) By symmetry between $P$ and $S$,
(\ref{pythagoreanPropositionEquality}) holds when $P$ is replaced by
$S$.

\vspace*{.1in}
\begin{proof}
1) $\Rightarrow$: Since $L_{\alpha}(P,R)$ and $L_{\alpha}(Q,R)$ are
finite, from (\ref{IDefinitionRho}), we gather that both $\sum_y
\sum_x P(x,y) R'(x \mid y)^{-\rho}$ and $\sum_y \sum_x Q(x,y) R'(x
\mid y)^{-\rho}$ are finite and nonzero.

Observe that $P_0 = Q$, and $L_{\alpha}\left( P_{\lambda}, R \right)
\geq L_{\alpha}\left( P_0, R \right)$ implies that
\[
I(P_{\lambda}, R) \geq I(P_0 , R).
\]
Thus
\begin{equation}
\label{limitingIfDerivative} \frac{I(P_{\lambda} , R) - I(P_0 ,
R)}{\lambda} \geq 0
\end{equation}
for every $\lambda \in (0,1]$. The limiting value as $\lambda
\downarrow 0$, the derivative of $I(P_{\lambda}, R)$ with respect to
$\lambda$ evaluated at $\lambda = 0$, should be $ \geq 0$. This will
give us the necessary condition.

Note that the derivative evaluated at $\lambda = 0$ is a one-sided
limit since $\lambda \in [0,1]$. We will first check that this
one-sided limit exists.

From (\ref{IDefinitionRho}), $I(P_{\lambda} , R)$ can be written as
$s(\lambda)/t(\lambda)$, where $t(\lambda)$ is bounded, positive,
and lower bounded away from 0, for every $\lambda$. Let $\dot{s}(0)$
and $\dot{t}(0)$ be the derivatives of $s$ and $t$ evaluated at
$\lambda = 0$. Clearly,
\begin{eqnarray*}
\dot{s}(0) & = & \lim_{\lambda \downarrow 0} \frac{s(\lambda) -
s(0)}{\lambda} \\
& = & \mbox{sign}(\rho) \left(\sum_y \sum_x P(x,y) \left( R'(x \mid
y) \right)^{-\rho} \right. \\
&& ~~~~~~~~~~~ \left. - \sum_y \sum_x Q(x,y) \left( R'(x \mid y)
\right)^{-\rho}\right).
\end{eqnarray*}
Similarly, it is easy to check that
\[
\dot{t}(0) = \sum_y \sum_x P(x,y) \left( Q'(x \mid y)
\right)^{-\rho} - t(0),
\]
with the possibility that it is $+ \infty$ (only when $0 < \alpha <
1$ and $\mbox{supp}(P) \not\subset \mbox{supp}(Q)$).

Since we can write
\begin{eqnarray*}
\lefteqn{ \frac{1}{\lambda}\left( \frac{s(\lambda)}{t(\lambda)} -
\frac{s(0)}{t(0)} \right) } \\
& = & \frac{1}{t(\lambda) t(0)} \left[ t(0) \frac{s(\lambda) -
s(0)}{\lambda} - s(0) \frac{t(\lambda) - t(0)}{\lambda} \right],
\end{eqnarray*}
it follows that the derivative of $s(\lambda)/t(\lambda)$ exists at
$\lambda =0$ and is given by $\left( t(0)\dot{s}(0) - s(0)\dot{t}(0)
\right) / t^2(0)$, with the possibility that it might be $+\infty$.
However, (\ref{limitingIfDerivative}) and $t(0) >0$ imply that
\[
\dot{s}(0) - s(0)\frac{\dot{t}(0)}{t(0)} \geq 0.
\]
Consequently, $\dot{t}(0)$ is necessarily finite. In particular,
when $0 < \alpha < 1$, we have ascertained that $L_{\alpha}(P,Q)$ is
finite. After substitution of $s(0), t(0), \dot{s}(0)$, and
$\dot{t}(0)$ we get
\begin{eqnarray}
\lefteqn{ \mbox{sign}(\rho) \cdot \sum_y \sum_{x} P(x,y) \left( R'(x \mid y) \right)^{-\rho}} \nonumber \\
& \geq & \mbox{sign}(\rho) \cdot \left( \sum_y \sum_{x} P(x,y)
\left(Q'(x
\mid y) \right)^{-\rho} \right) \nonumber \\
&& ~~~~~~~~~~~~ \cdot \left( \frac{\sum_y \sum_{x}
Q(x,y) \left(R'(x \mid y ) \right)^{-\rho}}{h(Q)} \right) \nonumber \\
\label{simplifiedPythagoreanInequality}
\end{eqnarray}
When $-1 < \rho < 0$, clearly, $\sum_y \sum_{x} P(x,y) \left(Q'(x
\mid y) \right)^{-\rho}$ cannot be zero, due to the nonzero
assumptions on the other quantities in
(\ref{simplifiedPythagoreanInequality}). This implies that
$L_{\alpha}(P, Q)$ is finite when $1 < \alpha < \infty$ as well. An
application of (\ref{LIrelationGeneral}) and (\ref{IDefinitionRho})
shows that (\ref{simplifiedPythagoreanInequality}) and
(\ref{pythagoreanPropositionInequality}) are equivalent. This
concludes the proof of the forward implication.

The reader will recognize that the basic idea is quite simple:
evaluation of a derivative at $\lambda = 0$ and a check that it is
nonnegative. The technical details above ensure that the case when
the derivative of the denominator is infinite is carefully examined.

1) $\Leftarrow$: The hypotheses imply that $L_{\alpha}(P,R),
L_{\alpha}(Q,R)$, and $L_{\alpha}(P,Q)$ are finite. As observed
above, (\ref{simplifiedPythagoreanInequality}) and
(\ref{pythagoreanPropositionInequality}) are equivalent. Observe
that both sides of (\ref{simplifiedPythagoreanInequality}) are
linear in $P$. This property will be exploited in the proof.
Clearly, if we set $P = Q$ in
(\ref{pythagoreanPropositionInequality}) and
(\ref{simplifiedPythagoreanInequality}), we have the equalities
\begin{equation}
\label{trivialPythagoreanEquality} L_{\alpha}(Q,R) = L_{\alpha}(Q,Q)
+ L_{\alpha}(Q,R)
\end{equation}
and
\begin{eqnarray}
\lefteqn{ \mbox{sign}(\rho) \cdot \sum_y \sum_{x} Q(x,y) \left( R'(x \mid y) \right)^{-\rho}} \nonumber \\
& = & \mbox{sign}(\rho) \cdot \left( \sum_y \sum_{x} Q(x,y)
\left(Q'(x
\mid y) \right)^{-\rho} \right) \nonumber \\
&& ~~~~~~~~~~~~ \cdot \left( \frac{\sum_y \sum_{x}
Q(x,y) \left(R'(x \mid y ) \right)^{-\rho}}{h(Q)} \right) \nonumber \\
\label{trivialPythagoreanEqualityExpanded}
\end{eqnarray}
A $\lambda$-weighted linear combination of the inequalities
(\ref{simplifiedPythagoreanInequality}) and
(\ref{trivialPythagoreanEqualityExpanded}) yields
(\ref{simplifiedPythagoreanInequality}) with $P$ replaced by
$P_{\lambda}$. The equivalence of
(\ref{pythagoreanPropositionInequality}) and
(\ref{simplifiedPythagoreanInequality}) result in
\begin{eqnarray*}
L_{\alpha}(P_{\lambda},R) & \geq & L_{\alpha}(P_{\lambda},Q) +
L_{\alpha}(Q,R) \\
& \geq & L_{\alpha}(Q,R).
\end{eqnarray*}
This concludes the proof of the first part.

2) This follows easily from the first statement. For the forward
implication, indeed, (\ref{simplifiedPythagoreanInequality}) holds
for $P$. Moreover, (\ref{simplifiedPythagoreanInequality}) holds
when $P$ is replaced by $S$. If either of these were a strict
inequality, the linear combination of these with the $\lambda$ given
by (\ref{innerPMF}) will satisfy
(\ref{trivialPythagoreanEqualityExpanded}) with strict inequality
replacing the equality, a contradiction. The reverse implication is
straightforward.
\end{proof}
\vspace*{.1in}

Let us now apply Proposition \ref{pythagoreanProposition} to the
$L_{\alpha}$-projection of a convex set. For a convex $\cal{E}$, we
call $Q$ an \emph{algebraic inner point of $\cal{E}$} if for every
$P \in {\cal{E}}$, there exist $S \in {\cal{E}}$ and $\lambda$
satisfying (\ref{innerPMF}).

\vspace*{.1in}
\begin{thm}{\em (Projection Theorem) }
\label{L-projection-theorem} Let $0 < \alpha < \infty, \alpha \neq
1$ and ${\mathbb{X}}$ a finite set. A PMF $Q \in {\cal{E}} \cap
B(R,\infty)$ is the $L_{\alpha}$-projection of $R$ on the convex set
${\cal{E}}$ if and only if every $P \in {\cal{E}}$ satisfies
\begin{equation}
\label{PythagoreanInequality} L_{\alpha}(P,R) \geq L_{\alpha}(P,Q) +
L_{\alpha}(Q,R).
\end{equation}

If the $L_{\alpha}$-projection $Q$ is an algebraic inner point of
${\cal{E}}$, then every $P \in {\cal{E}} \cap B(R,\infty)$
satisfies~(\ref{PythagoreanInequality}) with equality. $\hfill
\Box$.
\end{thm}
\vspace*{.1in}

\begin{proof}
This follows easily from Proposition \ref{pythagoreanProposition}.
For the case when $L_{\alpha}(P,R) = \infty$ not covered by
Proposition \ref{pythagoreanProposition},
(\ref{PythagoreanInequality}) holds trivially.
\end{proof}
\vspace*{.1in}

\begin{corollary}
Let $0 < \alpha <1$, and a PMF $Q \in {\cal{E}} \cap B(R,\infty)$ be
the $L_{\alpha}$-projection of $R$ on the convex set $\cal{E}$. If
$Q$ is an algebraic inner point of ${\cal{E}}$, then every $P \in
{\cal{E}}$ satisfies (\ref{PythagoreanInequality}) with equality.
\end{corollary}
\vspace*{.1in}

\begin{proof}
Clearly, for any $P \in {\cal{E}} $, we have $\mbox{supp}(P) \subset
\mbox{supp}(Q) \subset \mbox{supp}(R)$, and therefore ${\cal{E}}
\subset B(R,\infty)$. The corollary now follows from the second
statement of Theorem \ref{L-projection-theorem}.
\end{proof}
\vspace*{.1in}

While existence of $L_{\alpha}$-projection is guaranteed for certain
sets by Proposition \ref{existenceOfProjection}, the following talks
about uniqueness of the projection.

\vspace*{.1in}
\begin{proposition} ({\it Uniqueness of projection})
\label{uniquenessOfProjection} Let $0 < \alpha < \infty, \alpha \neq
1$. If the $L_{\alpha}$-projection of $R$ on the convex set
${\cal{E}}$ exists, it is unique.
\end{proposition}
\vspace*{.1in}
\begin{proof}
Let $Q_1$ and $Q_2$ be the projections. Then
\[
\infty > L_{\alpha}(Q_1, R) = L_{\alpha}(Q_2, R) \geq
L_{\alpha}(Q_2, Q_1) + L_{\alpha}(Q_1, R),
\]
where the last inequality follows from Theorem
\ref{L-projection-theorem}. Thus $L_{\alpha}(Q_2, Q_1) = 0$, and
$Q_2 = Q_1$.
\end{proof}
\vspace*{.1in}

Analogous to the Kullback-Leibler divergence case, our next result
is the transitivity property.

\vspace*{.1in}
\begin{thm}
Let $\cal{E}$ and ${\cal{E}}_1 \subset {\cal{E}}$ be convex sets of
PMFs on $\mathbb{X}$. Let $R$ have $L_{\alpha}$-projection $Q$ on
$\cal{E}$ and $Q_1$ on ${\cal{E}}_1$, and suppose that
(\ref{PythagoreanInequality}) holds with equality for every $P \in
{\cal{E}}$. Then $Q_1$ is the $L_{\alpha}$-projection of $Q$ on
${\cal{E}}_1$.
\end{thm}
\vspace*{.1in}
\begin{proof}
The proof is the same as in \cite[Theorem 2.3]{Csiszar-I-proj}. We
repeat it here for completeness.

Observe that from the equality hypothesis applied to $Q_1 \in
{\cal{E}}_1 \subset {\cal{E}}$, we have
\begin{equation}
\label{c1} L_{\alpha}(Q_1, R) = L_{\alpha}(Q_1, Q) + L_{\alpha}(Q,
R).
\end{equation}
Consequently $L_{\alpha}(Q_1, Q)$ is finite.

Furthermore, for a $P \in {\cal{E}}_1$, we have
\begin{eqnarray}
\lefteqn{L_{\alpha}(P, R)} \nonumber \\
\label{c2}  & \geq & L_{\alpha}(P, Q_1) + L_{\alpha}(Q_1, R) \\
\label{c3} & = & L_{\alpha}(P, Q_1) + L_{\alpha}(Q_1, Q) +
L_{\alpha}(Q, R),
\end{eqnarray}
where (\ref{c2}) follows from Theorem \ref{L-projection-theorem}
applied to ${\cal{E}}_1$, and (\ref{c3}) follows from (\ref{c1}).

We next compare (\ref{c3}) with $L_{\alpha}(P, R) = L_{\alpha}(P, Q)
+ L_{\alpha}(Q, R)$ and cancel $L_{\alpha}(Q, R)$ to obtain
\[
L_{\alpha}(P, Q) \geq L_{\alpha}(P, Q_1) + L_{\alpha}(Q_1, Q)
\]
for every $P \in {\cal{E}}_1$. Theorem \ref{L-projection-theorem}
guarantees that $Q_1$ is the $L_{\alpha}$-projection of $Q$ on
${\cal{E}}_1$.
\end{proof}
\vspace*{.1in}

As an application of Theorem \ref{L-projection-theorem} let us
characterize the $L_{\alpha}$-center of a family.

\vspace*{.1in}
\begin{proposition}
If the $L_{\alpha}$-center of a family $\mathbb{T}$ of PMFs exists,
it lies in the closure of the convex hull of the family.
\end{proposition}
\vspace*{.1in}

\begin{proof}
Let $\cal{E}$ be the closure of the convex hull of $\mathbb{T}$. Let
$Q^*$ be an $L_{\alpha}$-center of the family, and $C$, which is at
most $\log |\mathbb{X}|$, the $L_{\alpha}$-radius. Our first goal is
to show that $Q^* \in \cal{E}$.

By Proposition \ref{existenceOfProjection}, $Q^*$ has an
$L_{\alpha}$-projection $Q$ on $\cal{E}$, and by Proposition
\ref{uniquenessOfProjection}, the projection is unique on $\cal{E}$.
From Theorem \ref{L-projection-theorem}, for every $P \in
\mathbb{T}$, we have
\[
L_{\alpha}(P, Q^*) \geq L_{\alpha}(P, Q) + L_{\alpha}(Q, Q^*).
\]
Thus
\begin{eqnarray*}
C & = & \sup_{P \in \mathbb{T}} L_{\alpha}(P, Q^*) \\
& \geq & \sup_{P \in \mathbb{T}} L_{\alpha}(P, Q) + L_{\alpha}(Q, Q^*) \\
& \geq & C + L_{\alpha}(Q, Q^*).
\end{eqnarray*}
Thus $L_{\alpha}(Q, Q^*) = 0$, leading to $Q^* = Q \in {\cal{E}}$.
\end{proof}
\vspace*{.1in}

For the special case when $|\mathbb{T}| = m$ is finite, {\it i.e.},
$\mathbb{T} = \{ P_1, \cdots, P_m \}$, we found the weight vector
$w$ such that $Q^* = \sum_{i=1}^{m} w(i) P_{i}$ and $\sum_{i=1}^m
w(\theta) = 1$. This was done in an explicit fashion in Section
\ref{subsubsec:minimizerInConvexHull} using results on
$f$-divergences.

\section{Concluding remarks}

\label{conclusions}

We conclude this paper by applying some of our results to guessing
of strings of length $n$ with letters in $\mathbb{A}$. Let
$\mathbb{X} = \mathbb{A}^n$, $m = |\mathbb{A}|$, and $P$ a PMF on
$\mathbb{A}$. Let
\[
P_n(x^n) = \prod_{i=1}^n P(X_i = x_i)
\]
denote the PMF of the discrete memoryless source (DMS) where the
$n$-string $x^n = (x_1, x_2, \cdots, x_n)$. Theorem
\ref{ArikanGuessingTheorem} says that for $\rho = 1$, the minimum
expected number of guesses grows exponentially with $n$; the growth
rate is given by $H_{1/2}(P)$.

If the only information that the guesser has about the source is
that $P_n \in \mathbb{T}$, the guesser suffers a penalty
(interchangeably called redundancy); growth rate of the minimum
expected number of guesses is larger than that achievable with
knowledge of $P_n$. The increase in growth rate is given by the
normalized redundancy $R(P_n, G)/n$, where $G$ is the guessing
strategy chosen to work for all sources in $\mathbb{T}$. This
normalized redundancy equals the normalized $L_{1/2}$-radius of
$\mathbb{T}$, {\em i.e.}, $C_n/n$, where $C_n$ is given by
(\ref{minSupLalphaEqn}), to within $\log (1+n \ln m)$.

When $P_n$ is a DMS, and the PMF $P$ on $\mathbb{A}$ is unknown to
the guesser, Arikan and Merhav \cite{Arikan-Merhav} have shown that
guessing strings in the increasing order of their empirical
entropies is a universal strategy. Their universality result is
implied by the fact that the normalized $L_{1/2}$-radius of the
family of DMSs satisfies $C_n/n \rightarrow 0$. The family of DMSs
is thus not rich enough from the point of view of guessing.
Knowledge of the PMF $P$ is not needed; the universal strategy
achieves, asymptotically, the minimum growth rate achievable with
full knowledge of the source statistics.

Suppose now that $\mathbb{A} = \{0,1\}$; we may think of an
$n$-string as the outcome of independent coin tosses. Suppose
further that two biased coins are available. To generate each $X_i$,
one of the two coins is chosen arbitrarily, and tossed. The outcome
of the toss determines $X_i$. This is a two-state arbitrarily
varying source. We may assume $\mathbb{S} = \{ a, b \}$. Let us
assume that as $n \rightarrow \infty$, the fraction of time when the
first coin is picked approaches a limit $U^*(a)$. Let us further
assume that for each $n$, the receiver knows how many times the
first coin was picked, {\em i.e.}, it knows the type of the state
sequence. If the two coins are not statistically identical, the
normalized $L_{1/2}$-radius approaches a strictly positive constant
as $n \rightarrow \infty$. This implies that the growth rate in the
minimum expected number of guesses for a strategy without full
knowledge of source statistics is strictly larger than that
achievable with full knowledge of source statistics. We note that in
order to maximize the expected number of guesses, the right solution
may be to pick one coin, the one with the higher entropy, all the
time.

The guesser's lack of knowledge of the number of times the first
coin is picked results in additional redundancy. However this
additional redundancy asymptotically vanishes. The guesser
``stitches'' together the best guessing lists for each type of state
sequences.

%
%
%

\bibliographystyle{../sty/IEEEtran}
{
\bibliography{../sty/IEEEabrv,../wislBib/wisl}
}



%
%
%

\end{document}